\documentclass[prd,showpacs,preprintnumbers,showkeys,amsmath,amssymb]{revtex4}

\usepackage{amsmath,amssymb,amsthm}
\usepackage{color}
\usepackage{graphicx} 
\usepackage{epstopdf}
\usepackage{multirow}
\usepackage{subfigure}
\usepackage{rotating}
\usepackage{hyperref}
\usepackage{subfigure}

\graphicspath {{figure/}}
\usepackage{float}

\begin{document}

\title{Thermodynamical phase structures and particle emission rate of charged AdS black hole surrounded by string cloud and quintessence via shadow formalism}
\author{Yunxiang Wang, Hongyu Chen, Juhua Chen\footnote
{Corresponding author, Email: jhchen@hunnu.edu.cn}, Yongjiu Wang}
\affiliation{Department of Physics, Key Laboratory of Low Dimensional
 Quantum Structures and Quantum Control of Ministry of Education,
 and Synergetic Innovation Center for Quantum Effects and Applications,
  Hunan Normal University, Changsha, Hunan 410081, P. R. China.}

\begin{abstract}
In this paper, the shadow of the four-dimensional charged AdS black hole surrounded by string cloud and quintessence is derived. The shadow radius shows a strictly monotonic and invertible correlation with the event horizon radius. The phase structures of the black hole for different parameters are reproduced through traditional thermodynamic geometry, which are similar to a van der Waals system. By analyzing the phase structure of the black hole in the context of shadows, thermodynamical phase structures with the shadow radius as the variable replicate the phase transition with the event horizon radius as the variable. We present the energy emission rates for massless and massive particles and discover that the maximum emission frequency can also serve as a useful tool for thermodynamic analysis. We firstly study and systematically establish shadow thermodynamics under the background of string cloud and quintessence, and our results reveal the independent regulatory mechanism of dark components on phase transitions as well as the universal topological invariance of the phase transition structure.

\end{abstract}

\pacs{04.70.–s, 04.70.Dy, 04.50.Kd}
\keywords{shadow radius; phase transition; emission rate }

\maketitle
\section{Introduction}
\label{intro}
The black hole has always been a hot topic of study in gravitational physics and astrophysics. For example, in 2015, the Laser Interferometer Gravitational-Wave Observatory (LIGO) announced its first detection of gravitational waves, which came from the merger of a binary black hole\cite{abbott2016prl3,abbott2016prl2,abbott2016prl1}. The Event Horizon Telescope (EHT) captured images of the black hole at the center of the giant elliptical galaxy M87
\cite{AK2019L1} and the supermassive black hole in Sagittarius A*\cite{akiyama2022ajl12}. The strong gravitational field of a black hole distorts the spacetime around it and causes photons flying around the black hole to be captured in an unstable circular orbit\cite{Vazquez2004Lensing,Shaikh2019Shadows,Hou2018Shadow,Cunha2018Shadow,Tsukamoto2018Shadow}. The orbit formed by photons is called the photon sphere which means photons outside the critical orbit can move to infinity, while photons trapped in the critical orbit will gradually fall into the interior of the black hole. A static observer at infinity cannot observe these light rays, which naturally forms a dark region. The dark region is called the black hole shadow\cite{Wald1984GeneralRelativity}. The shape of black hole shadows has been widely studied, and rotating black holes can exhibit non-standard circular shapes because of different rotation parameters\cite{Bardeen1973,Chandrasekhar1983}. So far, black hole shadows have become a popular research topic\cite{Bambi2009,Bambi2010,Atamurotov2013,Papnoi2014,Atamurotov2015,Wang2018,Guo2018,Yan2019,Konoplya2019}.

The black hole shadow is an observable phenomenon which gives us information about the interior structure of the black hole and explains many observational effects. The photon sphere of the Schwarzschild black hole is located at r=3M\cite{Perlick2022,Synge1966b}, and Synge analyzed photon escape from Schwarzschild black holes and found that only photons emitted within a narrow cone around the radial direction can escape. The cone closes completely at the horizon, a result that directly defines the boundary of the black hole shadow\cite{Amarilla2010}. Luminet calculated the first realistic image of a black hole surrounded by a thin accretion disk through ray-tracing in Schwarzschild geometry. He showed that relativistic effects, especially gravitational lensing and Doppler shift, produce a highly asymmetric image: the receding side of the disk appears dimmer, while the approaching side appears brighter\cite{Luminet1979b}. Since the observation results were given, a large number of papers on black hole shadows have been provided to elucidate some observation results\cite{Afrin2021,Khodadi2021,Ghosh2022,Cunha2022}.

Hawking proved that black holes can radiate thermal particles\cite{Hawking1975,HawkingPage1983,Hawking1974}, which shows that black holes have temperature and entropy. Bekenstein proposed that black hole entropy is directly proportional to the area of its event horizon\cite{Bekenstein1973a,Bekenstein1972,Bekenstein1973b,Bekenstein1974,Bekenstein1975}. The groundbreaking idea drew an analogy between the laws of black hole mechanics and classical thermodynamics, and indicated that the information content of a black hole is encoded on its surface area. After the four laws of black hole thermodynamics were established\cite{Bardeen1973FourLaws}, black hole thermodynamics has been developing rapidly. In their extended phase space theory, the cosmological constant is understood as thermodynamic pressure. For the framework, a phase structure similar to a van der Waals system appears in black hole thermodynamics, and it has been found that the black hole state parameter has a close relationship with the van der Waals state equation, which generates critical behavior in AdS spacetime backgrounds\cite{Kubiznak2012,Cai2013,Kastor2009,Dolan2011}. Phase transition behavior is also studied in various black holes, such as topological and rotating black holes, which has led to a new field called black hole chemistry.

Hawking radiation is the core theory that connects general relativity, quantum field theory, and thermodynamics. Hawking proposed that quantum fluctuations occur near the black hole event horizon, which broke the traditional idea that the region around a black hole is a vacuum. The vacuum fluctuations near the event horizon will produce a pair of virtual particle pairs, where negative energy particles can pass through the event horizon and enter the black hole through the tunneling effect, while positive energy particles will be emitted outside the event horizon. Page\cite{Page1976} combined quantum field theory with perturbation theory to calculate the massless particle emission rate of the Schwarzschild black hole, and the calculation was later extended to other types of black holes. Emparan et al.\cite{Emparan2000} found that the Hawking radiation energy of a black hole in extra dimensions is mainly concentrated in four-dimensional Standard Model particles on the brane. 

There are many alternative theories of gravitational modification in general relativity, such as rainbow gravity\cite{Magueijo2004} and massive gravity\cite{deRham2010,deRham2011}. The research on black holes surrounded by string cloud was pioneered by Letelier\cite{Letelier1979}, whose core idea treats a collection of extended objects, such as one-dimensional strings\cite{Letelier1979}, as an additional source of the gravitational field, which generalizes the Schwarzschild solution and forms a spherically symmetric black hole model surrounded by string cloud. Astronomical observations confirm that the universe is in a state of accelerated expansion\cite{Perlmutter1999}, which means a repulsive effect exists in the universe pushing everything apart. Quintessence is a candidate for the repulsive effect, and Kiselev\cite{Kiselev2003} gave the black hole solution surrounded by quintessence. The thermodynamics of the AdS black hole has been widely studied because of the AdS/CFT correspondence\cite{Maldacena1998,Witten1998,Gubser1998,Cvetic2011,Chamblin1999a,Chamblin1999b,Caldarelli2000}. In this paper we will concentrate on the thermodynamics and particle emission rate of a charged AdS black hole in the background of string cloud and quintessence by using a novel method,  i.e. its maximum emission frequency and shadow radius. 

This paper is organized as follows. In Sec. \ref{RN dS structure}, we briefly review the horizon structure of the four-dimensional charged AdS black hole surrounded by string cloud and quintessence and derive its black hole shadow. In Sec. \ref{separated approach}, The phase structure of black hole thermodynamics will be studied via shadow formalism, and the size and shape of the black hole shadow will be given. In Sec. \ref{emission}, The energy emission rate around the black hole will be provided and linked to the structure of black hole thermodynamics. Finally, We provide a summary of these analysis results. (The units are $c=G=\hbar=k_{B}=1$)

\section{The shadow of the charged AdS black hole surrounded by string cloud and quintessence.}
\label{RN dS structure}
We consider a static spherically symmetric spacetime, which the line element is given by 
\begin{equation}
\begin{split}
&ds^2=-f(r)dt^2+\frac{1}{f(r)}dr^2+r^2(d\theta ^2+sin^2 \theta d\varphi^2).
\end{split}
\end{equation}
With this spacetime framework, the metric function\cite{Toledo2019} of the four‑dimensional charged AdS black hole surrounded by string cloud and quintessence is expressed as
\begin{equation}
f(r) =1 - a - \frac{2M}{r} + \frac{Q^2}{r^2} - \frac{\alpha}{r^{3\omega_q + 1}} - \frac{\Lambda r^2}{3},
\end{equation}
where $Q$ and $M$ are the charge and mass parameters of the black hole, $\Lambda$ is the cosmological constant in AdS spacetime, $a$ is the string cloud parameter, $\omega_q$ is the quintessence state parameter, and $\alpha$ is a positive normalization constant. In order to describe the scenario of an accelerating cosmic expansion, the quintessence state parameter is required to satisfy $-1 < \omega_q < -1/3$. In what follows, we shall fix $\omega_q = -2/3$.

In AdS spacetime, the negative curvature property can be interpreted as the thermodynamic pressure $P$ in the black hole thermodynamics framework\cite{Kastor2009}, which is related to the cosmological constant $\Lambda$ by
 \begin{equation}
 \begin{split}  
P = -\frac{\Lambda}{8\pi}
\end{split}
\end{equation}

To establish the relationship between the shadow radius and the event horizon radius for the black hole, we first consider the motion of photons around the black hole, which is governed by the Euler–Lagrange equations,
\begin{equation}
	\label{e4}
\begin{split}
\frac{\mathrm{d}}{\mathrm{d}\sigma}\left( \frac{\partial \mathcal{L}}{\partial \dot{x}^\mu} \right) = \frac{\partial \mathcal{L}}{\partial x^\mu},
\end{split}
\end{equation}
in which $\sigma$ is the affine parameter, $\dot{x}^\mu$ denotes the photon four-velocity, and $\mathcal{L}$ is the Lagrangian density of the photon system,
\begin{equation}
\begin{split}
\mathcal{L} = -\frac{1}{2} g_{\mu\nu} \frac{dx^\mu}{d\sigma} \frac{dx^\nu}{d\sigma} = \frac{1}{2}\left(f(r)\dot{t}^2 - \frac{\dot{r}^2}{f(r)} - r^2\left(\dot{\theta}^2 + \sin^2\theta \dot{\phi}^2\right)\right).
\end{split}
\end{equation}
The four-momentum of the photon is given by
\begin{equation}
\begin{split}
p_\mu = \frac{\partial \mathcal{L}}{\partial \dot{x}^\mu}.
\end{split}
\end{equation}
By confining the photon's motion to the equatorial plane with $\theta=\pi/2$ and $\dot{\theta}=0$ , the metric no longer depends on the coordinates $t$ and $\phi$, which gives rise to the corresponding conserved energy $E$ and angular momentum $L$\cite{Li2023},
\begin{equation}
\begin{split}
E = -g_{tt} \frac{dt}{d\sigma} = f(r) \frac{dt}{d\sigma},
\end{split}
\end{equation}
\begin{equation}
\begin{split}
 \quad L = g_{\phi\phi} \frac{d\phi}{d\sigma} = r^2 \frac{d\phi}{d\sigma}.
\end{split}
\end{equation}
The effective potential for the photon's radiative motion follows from the null-geodesic normalization condition 
$g_{\mu\nu}\dot{x}^{\mu}\dot{x}^{\nu}=0$, and satisfies the form,
\begin{equation}
\begin{split}
	\label{e9}
\dot{r}^2 + V_{\text{eff}}(r) = 0.
\end{split}
\end{equation}
The effective potential can be expressed as
\begin{equation}
\begin{split}
	\label{e10}
V_{\text{eff}}(r) = f(r)\left( \frac{L^2}{r^2} - \frac{E^2}{f(r)} \right).
\end{split}
\end{equation}
By combining Eq. (\ref{e9}) and Eq. (\ref{e10}), the orbital equation describing the photon’s motion is given by
\begin{equation}
\begin{split}
\frac{dr}{d\phi} = \pm r \sqrt{f(r)\left( \frac{r^2 E^2}{L^2 f(r)} - 1 \right)}.
\end{split}
\end{equation}
The critical photon orbit satisfies the following conditions on the effective potential,
\begin{equation}
\label{e12}
\begin{split}
V_{\text{eff}}(r_p) = 0, \quad \frac{dV_{\text{eff}}(r)}{dr}\bigg|_{r=r_p} = 0.
\end{split}
\end{equation}
Given the turning point of the photon’s orbital motion,
\begin{equation}
\begin{split}
\left. \frac{dr}{d\phi} \right|_{r=r_p} = 0,
\end{split}
\end{equation}
which yields the equation,
\begin{equation}
\begin{split}
\frac{dr}{d\phi} = \pm r \sqrt{f(r)\left( \frac{r^2 f(r_p)}{r_p^2 f(r)} - 1 \right)}.
\end{split}
\end{equation}

Following Refs.\cite{Zhang2020,Belhaj2020,Cai2021arXiv,Wang2022}, a static observer at position $r_{0}$ sends a light ray into the past, which forms an angle $\beta$ relative to the radical direction given by
\begin{equation}
\begin{split}
\cot \beta = \left. \frac{\sqrt{f_{rr}}}{\sqrt{f_{\phi\phi}}} \frac{dr}{d\phi} \right|_{r=r_0} .
\end{split}
\end{equation}
Applying elementary trigonometric relations, we obtain
\begin{equation}
\begin{split}
\sin^2 \beta = \frac{f(r_0) r_p^2}{r_0 f(r_p)}.
\end{split}
\end{equation}
Consequently, when $r$ tends to the photon sphere radius $r_p$, the shadow radius observed by a static observer situated at $r_0$ is expressed as
\begin{equation}
\begin{split}
\label{e17}
r_s = r_0 \sin \beta = r_p \sqrt{\frac{f(r_0)}{f(r_p)}}.
\end{split}
\end{equation}
According to Eq.(\ref{e4}) and Eq.(\ref{e12}), when $V_{\text{eff}}' = 0$, the photon sphere radius for the four-dimensional charged AdS black hole surrounded by quintessence and a string cloud is given by 
\begin{equation}
\begin{split}
(2-2a)r_p^2 - 6Mr_p + 4Q^2 - \alpha r_p^3 = 0.
\end{split}
\end{equation}
The photon sphere radius is determined by solving the constraint equation numerically. For a static observer located at spatial infinity, we have $f(r_{0})\to1$\cite{Zhang2020}, which allows us to determine the shadow radius.

\begin{figure}[htbp]
	\begin{minipage}{0.325\linewidth}
		\centerline{\includegraphics[height=5cm,width=5cm]{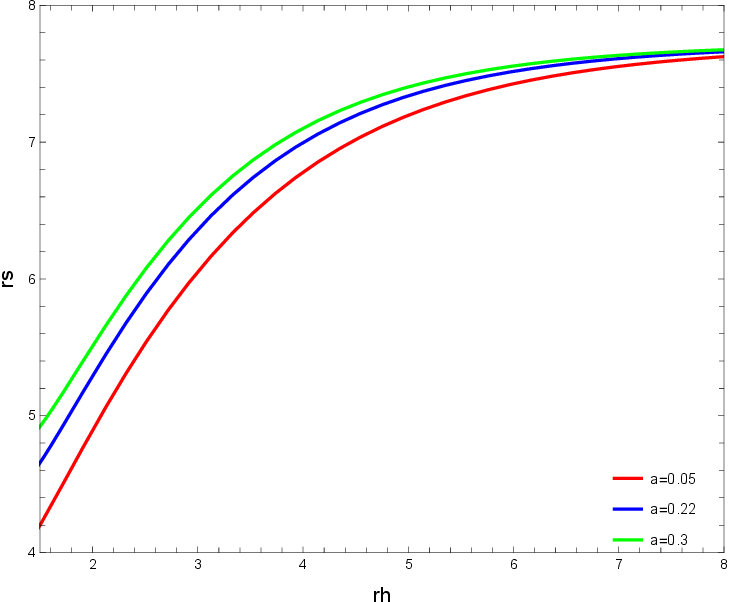}}
		\centerline{(a)}
	\end{minipage}
	\begin{minipage}{0.325\linewidth}
		\centerline{\includegraphics[height=5cm,width=5cm]{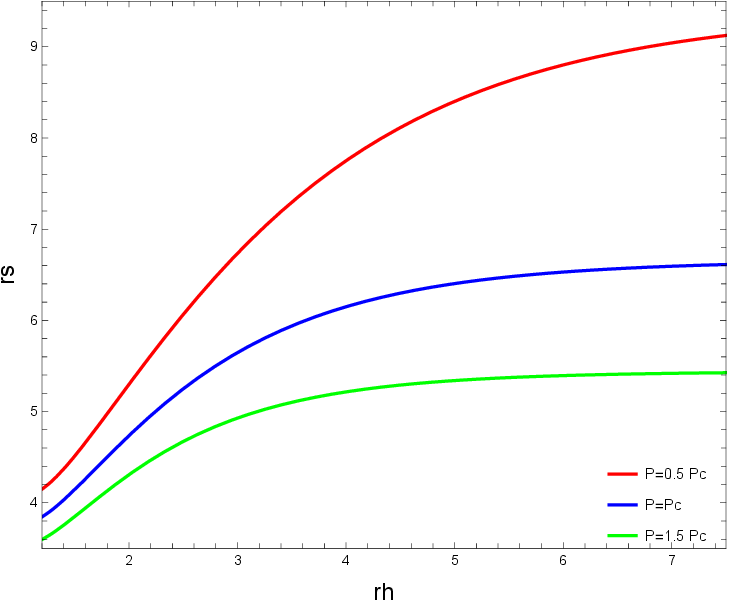}}
		\centerline{(b)}
	\end{minipage}
	\begin{minipage}{0.325\linewidth}
		\centerline{\includegraphics[height=5cm,width=5cm]{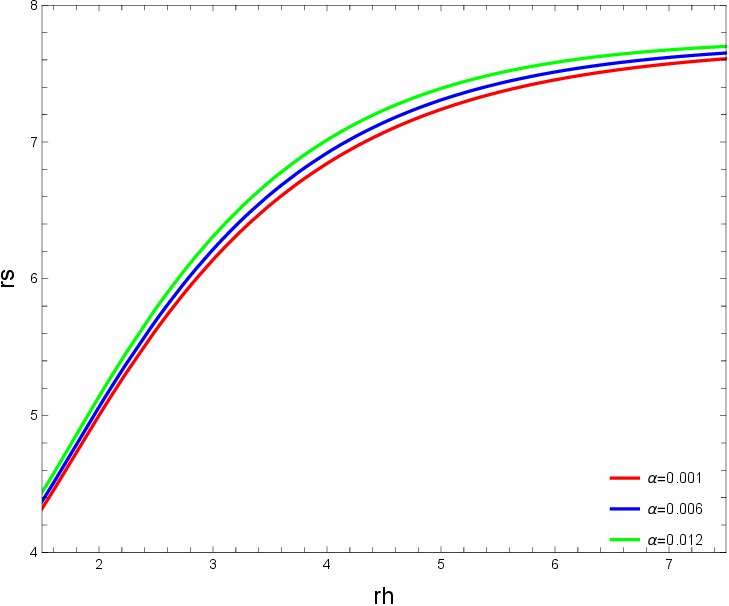}}
		\centerline{(c)}
	\end{minipage}
	\caption{The shadow radius $r_{s}$ as a function of the horizon radius $r_{h}$. Figure $(a)$: Fixed the thermodynamic pressure $P=0.002$ and the quintessence parameter $\alpha=0.001$, with varying string cloud parameter $a$. Figure $(b)$: Fixed the string cloud parameter $a=0.1$ and the quintessence parameter $\alpha=0.001$, with varying thermodynamic pressure $P$. Figure $(c)$: Fixed the thermodynamic pressure $P=0.002$ and the string cloud parameter $a=0.1$, with varying quintessence parameter $\alpha$. Here we set $Q=1, r_0=100$.} \label{Fig1}
\end{figure}

Figure \ref{Fig1} shows how the shadow radius $r_s$ changes with the horizon radius $r_h$. In all figures, $r_s$ grows as $r_h$ increases and gradually flattens out. In Figure (a), a larger string cloud parameter $a$ leads to a larger shadow radius, and all curves tend toward the same limiting value. Figure (b) shows that the shadow radius becomes smaller when the pressure $P$ is higher. Figure (c) shows the same trend as Figure (a), where the shadow radius increases with the quintessence parameter $\alpha$. Overall, $r_s$ increases as $r_h$ increases, and the positive correlation is independent of other parameters. The results give us an insight: we can use the quasi correlation between the shadow radius and the event horizon radius to probe physical behaviors of black holes, because while the event horizon radius is a traditional geometric quantity that is difficult to observe directly, the shadow radius is in fact an physical observable.

\section{Thermodynamics of the charged AdS black hole surrounded by string cloud and quintessence using shadow formalism.}
\label{separated approach}
In this section, we briefly review the thermodynamics of the four‑dimensional charged AdS black hole surrounded by string cloud and quintessence, and then build the connection between the shadow radius and the black hole thermodynamics. When $f(r) = 0$, we can obtain
\begin{equation}
\begin{split}
M=\frac{r_h}{2}\left(1-a+\frac{Q^2}{r_h^2}+\frac{8\pi P}{3}r_h^2-\alpha r_h\right).
\end{split}
\end{equation}
The Hawking temperature determined by the surface gravity is given by
\begin{equation}
	\label{e20}
\begin{split}
T=\frac{1}{4\pi}\left(\frac{1-a}{r_h}-\frac{Q^2}{r_h^3}+8\pi P r_h-2\alpha\right).
\end{split}
\end{equation}
The state equation can be expressed as
\begin{equation}
\begin{split}
P = \frac{4\pi T r_h^3 - (1-a) r_h^2 + q^2 + 2\alpha r_h^3}{8\pi r_h^4}.
\end{split}
\end{equation}
Next, we determine the thermodynamic critical point for the black hole by solving the following equations
\begin{equation}
	\label{e22}
\begin{split}
\left.\frac{\partial T}{\partial r_h}\right|_{r_c, P_c} = \left.\frac{\partial^2 T}{\partial r_h^2}\right|_{r_c, P_c} = 0.
\end{split}
\end{equation}
We can determine these critical points:
\begin{equation}
\begin{split}
r_c = \sqrt{\frac{6}{1-a}} Q, \quad P_c = \frac{(1-a)^2}{96\pi Q^2}, \quad T_c = \frac{\sqrt{6}\,(1-a)^{3/2} - 9\alpha Q}{18\pi Q}.
\end{split}
\end{equation}
Then, we can obtain the Gibbs free energy $G = M - TS$
\begin{equation}
\begin{split}
G = \frac{1}{4}\left(r_h(1 - a) + \frac{3Q^2}{r_h} - \frac{8\pi P r_h^3}{3}\right).
\end{split}
\end{equation}

\begin{figure}[htbp]
	\begin{minipage}{0.475\linewidth}
		\centerline{\includegraphics[height=4.8cm,width=6cm]{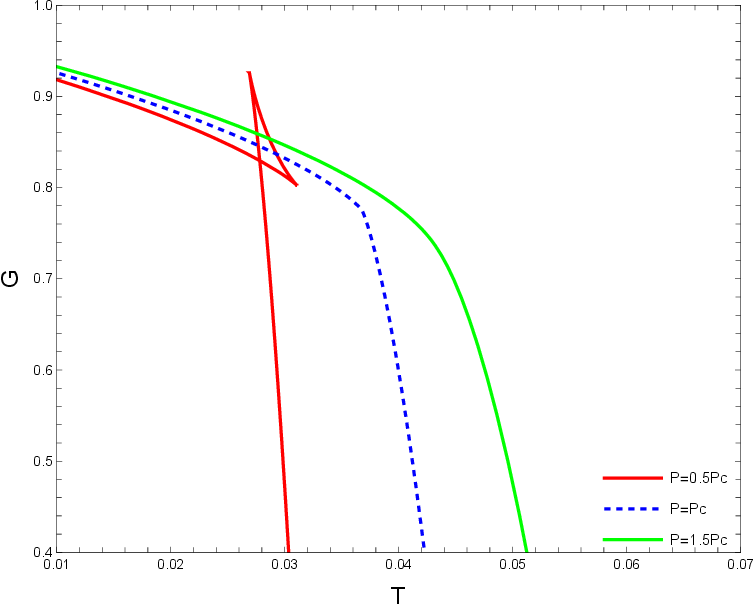}}
		\centerline{(a)}
	\end{minipage}
	\begin{minipage}{0.475\linewidth}
		\centerline{\includegraphics[height=4.8cm,width=6cm]{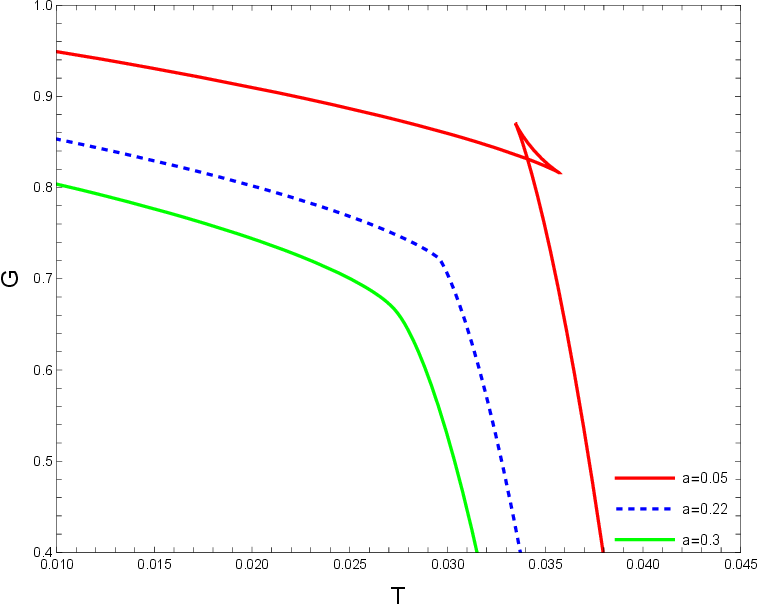}}
		\centerline{(b)}
	\end{minipage}
	\caption{The Gibbs free energy \( G \) as a function of the Hawking temperature \( T \). Figure $(a)$: Fixed the string cloud parameter \( a=0.1 \) and the quintessence parameter \( \alpha=0.001 \), with varying thermodynamic pressure \( P \). Figure $(b)$: Fixed the thermodynamic pressure \( P=0.002 \) and the quintessence parameter \( \alpha=0.001 \), with varying string cloud parameter \( a \). Here we set \( Q=1 \).}
	\label{Fig2}
\end{figure}

From Eq.(\ref{e22}), we can see that the temperature $T$ changes with the horizon radius $r_h$ when the parameters are fixed. From our analysis of the critical values, we observe that both the string cloud parameter $a$ and the thermodynamic pressure $P$ are related to phase transition behavior, while the quintessence parameter $\alpha$ only affects the magnitude of the temperature. Therefore, we now focus on analyzing the effects of the string cloud parameter $a$ and the thermodynamic pressure $P$ on the thermodynamic temperature $T$. Fig.\ref{Fig2}  shows the variation of the Gibbs free energy as a function of the temperature. For $P > P_c$, we find that no phase transition occurs in the $G$-$T$ diagram. When $P$ is below the critical point, the $G$-$T$ diagram develops a swallow tail shape, revealing a first order phase transition. Similar behavior is shown in the figure(b): The $G$-$T$ diagram shows no phase transition when the string cloud parameter $a>0.22$. Conversely, for $a<0.22$, it exhibits a swallow tail structure, indicating a phase transition. These phenomena must be related to the behavior of the Hawking temperature, so we next simulate the Hawking temperature $T$ as a function of the horizon radius $r_{h}$.

\begin{figure}[htbp]
\begin{minipage}{0.475\linewidth}
	\centerline{\includegraphics[height=4.8cm,width=6cm]{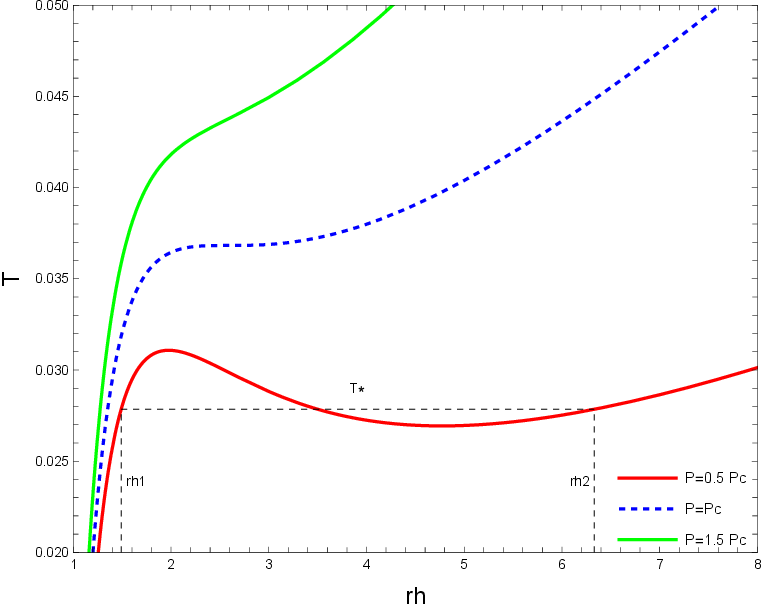}}
	\centerline{(a)}
\end{minipage}
\begin{minipage}{0.475\linewidth}
	\centerline{\includegraphics[height=4.8cm,width=6cm]{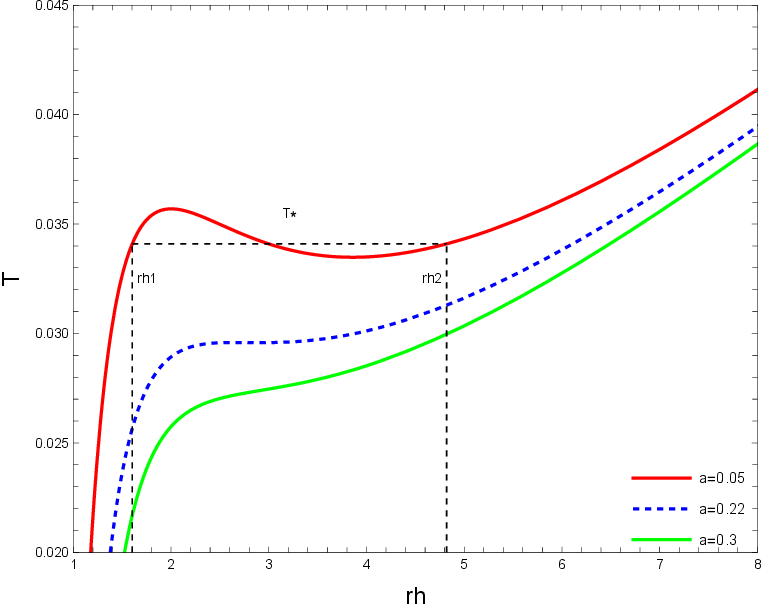}}
	\centerline{(b)}
\end{minipage}
\caption{The Hawking temperature \( T \) as a function of the horizon radius \( r_{h} \). Figure $(a)$: Fixed the string cloud parameter \( a=0.1 \) and the quintessence parameter \( \alpha=0.001 \), with varying thermodynamic pressure \( P \). Figure $(b)$: Fixed the thermodynamic pressure \( P=0.002 \) and the quintessence parameter \( \alpha=0.001 \), with varying string cloud parameter \( a \). $T^*$ (black dashed line) is the coexistence temperature. Here we set \( Q=1 \).}
\label{Fig3a}
\end{figure}

Based on Eq.(\ref{e20}), the $T$-$r_h$ diagrams for three different pressure values are presented in Fig.\ref{Fig3a} (a) , where the thermodynamic structure of the black hole exhibits distinct characteristics depending on the pressure. It can be seen that for $P > P_{c}$, the temperature curve is monotonic, with a slope that first decreases and then increases but never drops to zero, indicating that the system is in a supercritical phase without any phase transition. At $P = P_{c}$, the curve develops an inflection point, marking the critical isobaric where the slope vanishes and then recovers, which signals the beginning of critical behavior and indicates that the system is thermodynamically unstable at this point. When $P < P_{c}$, the temperature curve is no longer monotonic and instead shows the van der Waals-like phase transition behavior. In this regime, two stable branches emerge, which are connected by an unstable intermediate branch. One branch lies in the small radius region, corresponding to the fluid phase of the van der Waals system, and the other is in the large radius region, corresponding to the gas phase. According to Maxwell's law of equal areas\cite{Kubiznak2012},
\begin{equation}
\int_{r_{\rm h1}}^{r_{\rm h2}} T d r_{\rm h} = T^* (r_{\rm h2} - r_{\rm h1}),
\end{equation}
we can calculate the area between the small radius $r_{h1}$ and the large radius $r_{h2}$. The curve first increases, then decreases, and finally increases again, which also explains the swallowtail behavior of the $G$-$T$ diagram. For $r_{h}<r_{h1}$, the system corresponds to a stable small black hole. Once $r_{h}$ exceeds $r_{h2}$, it transitions into a stable large black hole. The intermediate interval $r_{h1}<r_{h}<r_{h2}$ is thermodynamically unstable and reflects the coexistence region where small and large black holes overlap during the phase transition. In Figure (b), the temperature $T$ follows a similar trend which is plotted against the string cloud parameter $a$. The key difference is that a larger $a$ suppresses the temperature, causing its overall value to drop.

Now we return our attention to the analysis of the shadow radius $r_{s}$ and the horizon radius $r_{h}$. A key physical question is whether the black hole shadow can still faithfully reflect the phase transition behavior under the double corrections from string cloud and quintessence. The present study answers the question and further reveals that different dark components have independent regulatory roles. The string cloud parameter governs the existence of the phase transition, while the quintessence parameter only adjusts the temperature amplitude. To make the analysis strictly valid, a proof of the monotonic relationship between $r_s$ and $r_h$ is provided. 

Based on Eq.(\ref{e17}), we can obtain
\begin{equation}
\label{e26}
	\left. \frac{dr_s}{dR} \right|_{R\to r_p} = \left\{ \left[ \frac{f(r_0)}{f(R)} \right]^{\frac{3}{2}} \frac{2f(r)-r_p f'(r)}{2f(r_0)} \right\}\bigg|_{R\to r_p} > 0.
\end{equation}
According to Eq.(\ref{e10}) and combining with the photon motion equation, we can obtain
\begin{equation}
	\left. \frac{df(R)}{dR} \right|_{R\to r_p} > 0.
\end{equation}
Since the Hawking temperature of a black hole must be positive, the following relation holds
\begin{equation}
	\left. \frac{df(r)}{dr} \right|_{r\to r_h} = \left. \frac{df(R)}{dR}\frac{dR}{dr_h} \right|_{R\to r_p} > 0.
\end{equation}
Simplifying the above equation yields
\begin{equation}
\label{e29}
	\left. \frac{dR}{dr_h} \right|_{R\to r_p} > 0.
\end{equation}
Considering Eqs.(\ref{e26}) and (\ref{e29}), using the chain rule
\begin{equation}
	\frac{dr_s}{dr_h} = \frac{dr_s}{dR}\frac{dR}{dr_h} > 0.
\end{equation}
For $R \to r_p$, the black hole temperature satisfies
\begin{equation}
	\frac{\partial T}{\partial r_h} = \frac{\partial T}{\partial r_s}\frac{\partial r_s}{\partial r_h},
\end{equation}
which indicates that the relation between the temperature $T$ and the horizon radius $r_h$ is in accordance with that between the temperature $T$ and the shadow radius $r_s$,
\begin{equation}
	\frac{\partial T}{\partial r_h} > 0,\quad \frac{\partial T}{\partial r_h} = 0,\quad \frac{\partial T}{\partial r_h} < 0,
\end{equation}
\begin{equation}
	\frac{\partial T}{\partial r_s} > 0,\quad \frac{\partial T}{\partial r_s} = 0,\quad \frac{\partial T}{\partial r_s} < 0.
\end{equation}
We have verified that the first order derivative of the shadow radius $r_s$ with respect to the horizon radius $r_h$ is positive via both theoretical and numerical analyses, which demonstrated  the thermodynamic phase structure of black holes can be characterized by the shadow radius. The monotonicity and reversibility are preserved even under the combined effects of string cloud, quintessence, and thermodynamic pressure, which demonstrates that the observable-based proxy method is strongly universal and not limited to simple black hole spacetimes.

\begin{figure}[htbp]
\begin{minipage}{0.475\linewidth}
	\centerline{\includegraphics[height=4.8cm,width=6cm]{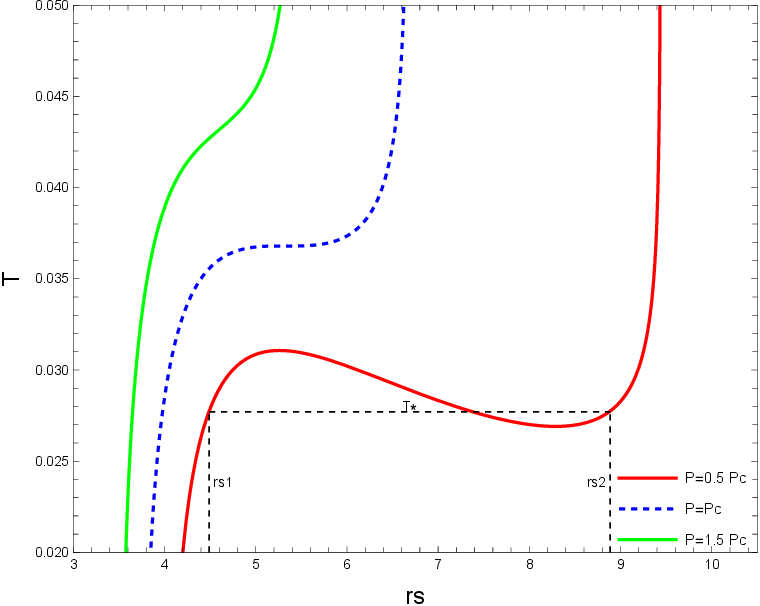}}
	\centerline{(a)}
\end{minipage}
\begin{minipage}{0.475\linewidth}
	\centerline{\includegraphics[height=4.8cm,width=6cm]{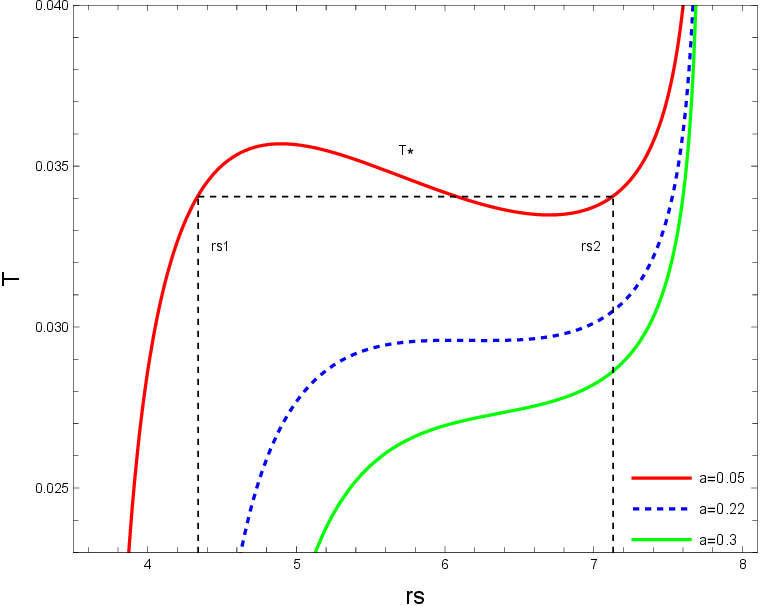}}
	\centerline{(b)}
\end{minipage}
\caption{The Hawking temperature $T$ as a function of the shadow radius $r_{s}$. Figure $(a)$: Fixed the string cloud parameter $a=0.1$ and the quintessence parameter $\alpha=0.001$, with varying thermodynamic pressure $P$. Figure $(b)$: Fixed the thermodynamic pressure $P=0.002$ and the quintessence parameter $\alpha=0.001$, with varying string cloud parameter $a$. $T^*$ (black dashed line) is the coexistence temperature. Here we set $Q=1, r_0=100$. }
\label{Fig4}
\end{figure}

 In the Fig.\ref{Fig4} , we show the Hawking temperature $T$ as a function of the shadow radius $r_{s}$ for different string cloud parameters and pressures. We observe a behavior that replicates the structure displayed in the $T$-$r_{h}$ plane by using the black hole shadow radius as the variable for the temperature $T$ and also reproduces the phase transition structure of the black hole. The corresponding unstable transition interval $r_{s1}<r_{s}<r_{s2}$ is marked by black dashed lines in the figure. These results indicate that black hole shadows can serve as an effective tool for investigating the thermodynamic behavior of black holes. The fact that the shadow radius can fully reproduce the phase transition structure means black hole thermodynamic phase transitions have a topological invariance. When the horizon radius is replaced by the shadow radius, the topological structure of the phase transition stays the same structure. Such structural invariance for different background fields shows an intrinsic property of black hole thermodynamics.

\begin{figure}[htbp]
	\begin{minipage}{0.475\linewidth}
		\centerline{\includegraphics[height=4.8cm,width=6cm]{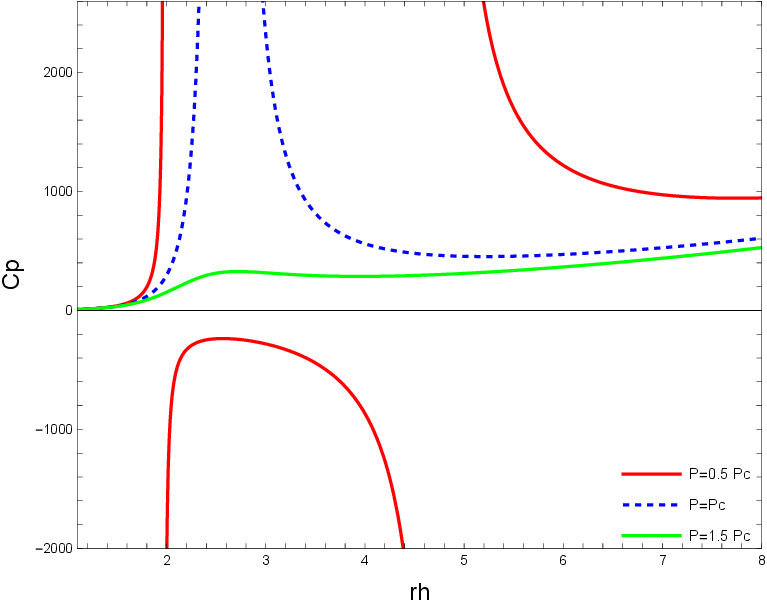}}
		\centerline{(a)}
	\end{minipage}
	\begin{minipage}{0.475\linewidth}
		\centerline{\includegraphics[height=4.8cm,width=6cm]{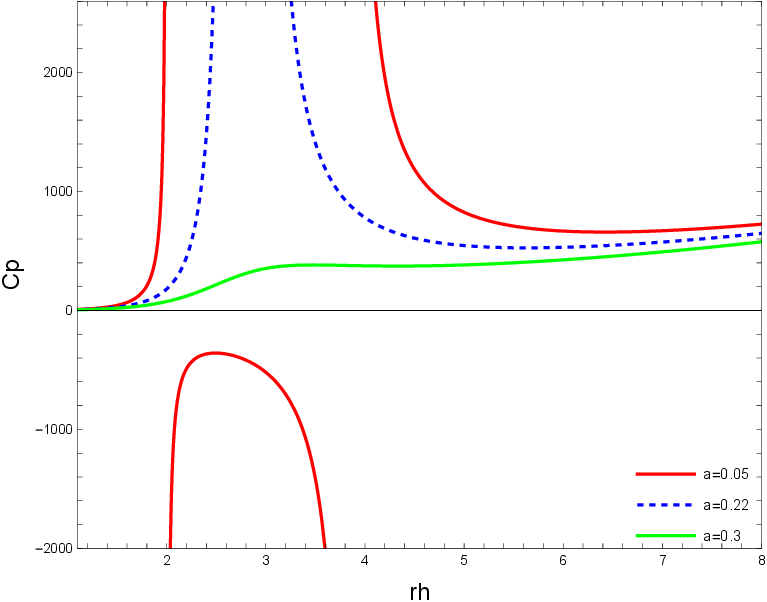}}
		\centerline{(b)}
	\end{minipage}
	\caption{The heat capacity $C_{P}$ as a function of the horizon radius $r_{h}$. Figure $(a)$: Fixed the string cloud parameter $a=0.1$ and the quintessence parameter $\alpha=0.001$, with varying thermodynamic pressure $P$. Figure $(b)$: Fixed the thermodynamic pressure $P=0.002$ and the quintessence parameter $\alpha=0.001$, with varying string cloud parameter $a$. Here we set $Q=1$. }
	\label{Fig5}
\end{figure}

\begin{figure}[htbp]
	\begin{minipage}{0.475\linewidth}
		\centerline{\includegraphics[height=4.8cm,width=6cm]{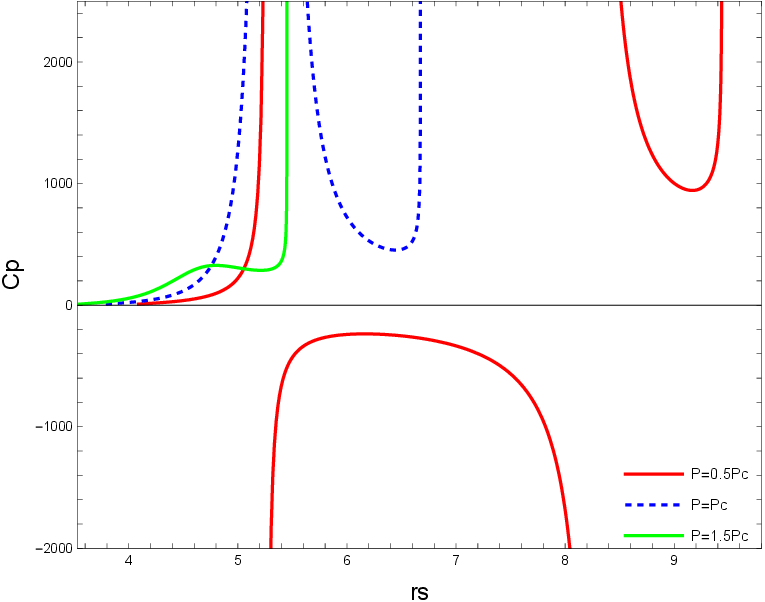}}
		\centerline{(a)}
	\end{minipage}
	\begin{minipage}{0.475\linewidth}
		\centerline{\includegraphics[height=4.8cm,width=6cm]{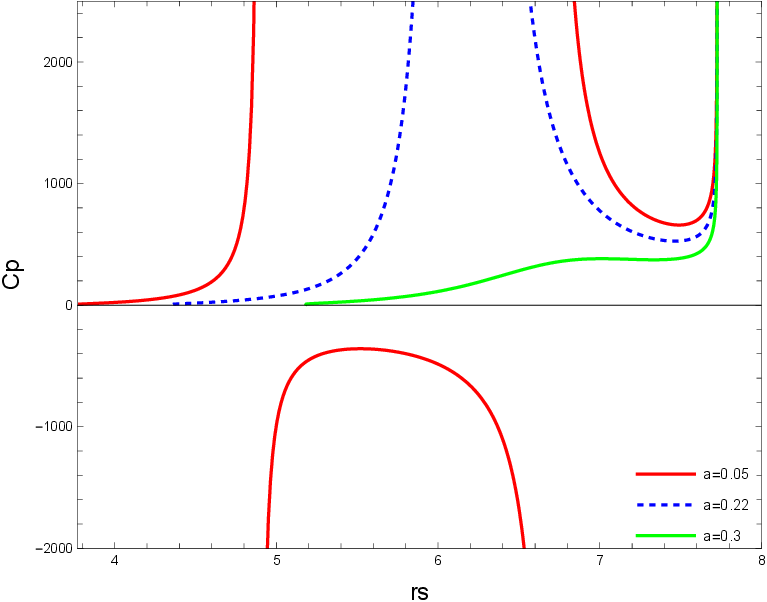}}
		\centerline{(b)}
	\end{minipage}
	\caption{The heat capacity $C_{P}$ as a function of the shadow radius $r_{s}$. Figure $(a)$: Fixed the string cloud parameter $a=0.1$ and the quintessence parameter $\alpha=0.001$, with varying thermodynamic pressure $P$. Figure $(b)$: Fixed the thermodynamic pressure $P=0.002$ and the quintessence parameter $\alpha=0.001$, with varying string cloud parameter $a$. Here we set $Q=1, r_0=100$. }
	\label{Fig6}
\end{figure}

The relationship diagram between black hole heat capacity and horizon radius is also a crucial method for investigating black hole thermal stability. The heat capacity $C_{P}$ is given by 
\begin{equation}
C_p = T \left. \frac{\partial S}{\partial T} \right|_p = \frac{2 \pi r_h^4 \left(8 \pi P r_h^2 + 1 - a - 2\alpha\right) + 2 \pi Q^2 r_h^2}{8 \pi P r_h^4 + r_h^2 (a - 1) + 3 Q^2}.
\end{equation}
and the heat capacity $C_{P}$ versus the horizon radius $r_{h}$ or the shadow radius $r_{s}$ in different cases is shown in Fig.\ref{Fig5} and Fig.\ref{Fig6} . In the $C_{P}$-$r_{h}$ diagram, the heat capacity diverges at the critical point, indicating the occurrence of a phase transition. When $P < P_c$, the heat capacity is negative and discontinuous at two separated points. Therefore, the system is thermodynamically unstable in the region bounded by these two divergence points, while outside this region it returns to thermodynamical stability. When $P > P_c$, the divergence points disappear, indicating that no phase transition occurs in the system. Following the shadow analysis method, we analyze the $C_{P}$-$r_{s}$ diagram using $r_{s}$ instead of $r_{h}$ again. The figure clearly shows the same results as in the $C_{P}$-$r_{h}$ plane, which reveals the complete phase transition structure. The correlation between the heat capacity and the shadow radius further strengthens our approach. In fact, the shadow radius itself is an observable, and compared to the horizon radius $r_{h}$, using it to reveal the thermodynamic structure is more advantageous.

\begin{figure}[htbp]
	\begin{minipage}{0.325\linewidth}
		\centerline{\includegraphics[height=5cm,width=5cm]{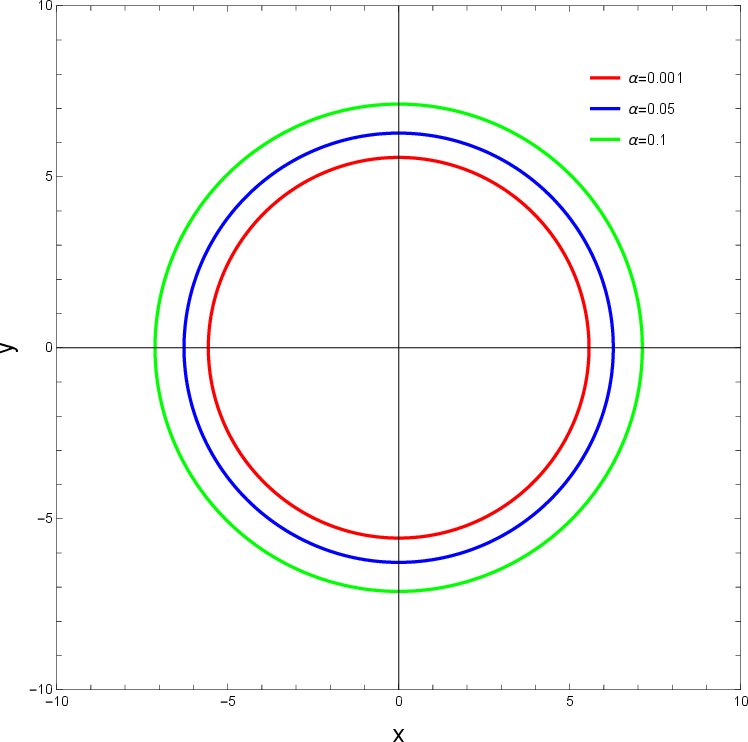}}
		\centerline{(a)}
	\end{minipage}
	\begin{minipage}{0.325\linewidth}
		\centerline{\includegraphics[height=5cm,width=5cm]{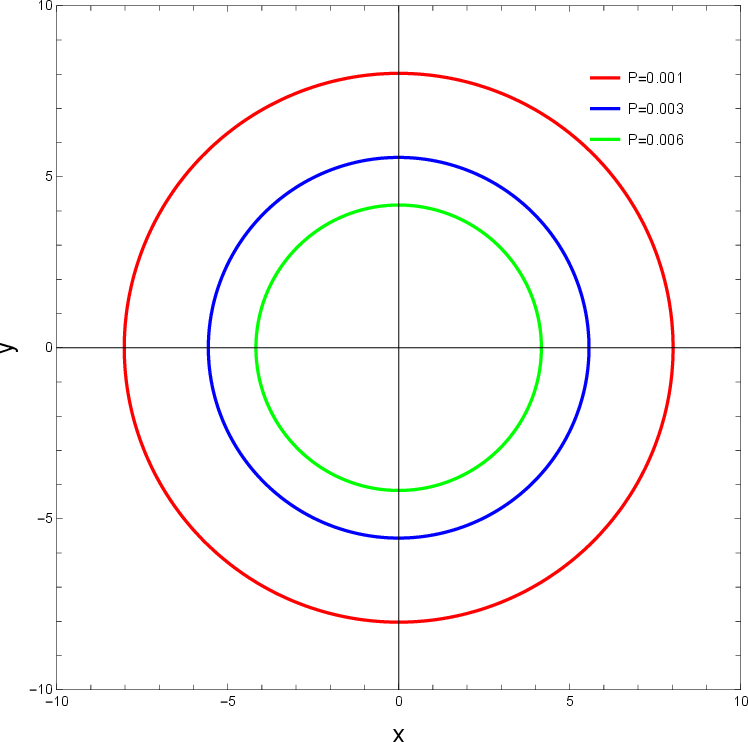}}
		\centerline{(b)}
	\end{minipage}
	\begin{minipage}{0.325\linewidth}
		\centerline{\includegraphics[height=5cm,width=5cm]{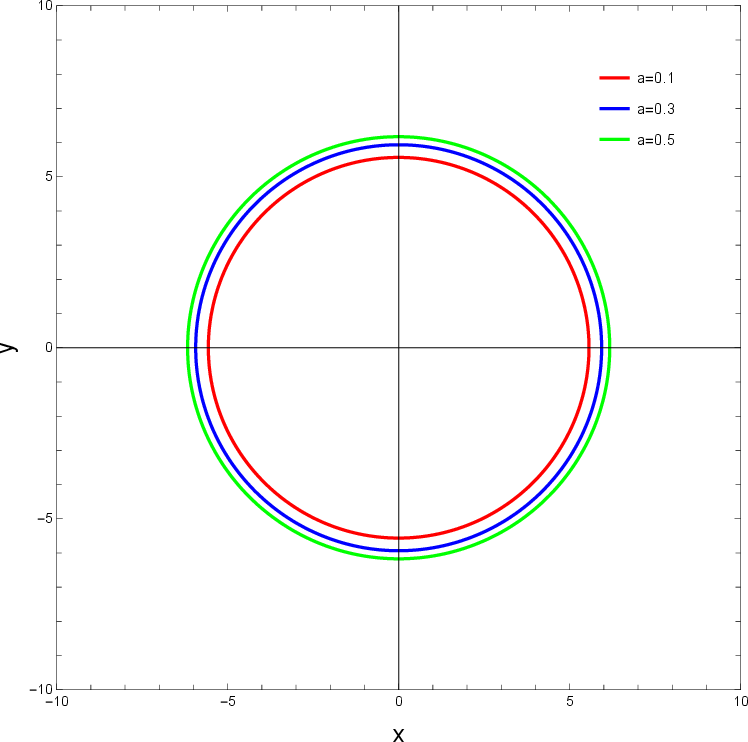}}
		\centerline{(c)}
	\end{minipage}
	\caption{The shadow shape of the four-dimensional charged AdS black hole surrounded by quintessence and string cloud for different parameters.
	Figure $(a)$: Fixed the string cloud parameter $a=0.1$ and the thermodynamic pressure $P=0.003$, with varying quintessence parameter $\alpha$.
	Figure $(b)$: Fixed the string cloud parameter $a=0.1$ and the quintessence parameter $\alpha=0.001$, with varying thermodynamic pressure $P$.
	Figure $(c)$: Fixed the quintessence parameter $\alpha=0.001$ and the thermodynamic pressure $P=0.003$, with varying string cloud parameter $a$.
	Here we set $Q=1, r_0=100, M=2$.} \label{Fig7}
\end{figure}

In the final part of this section, to visualize the results more clearly, we present how the size of the shadow radius changes with different parameters. In the vertical plane of the observer's line of sight, the shape of the shadow radius can be expressed in celestial coordinates. The celestial coordinates of an observer far from the black hole are defined as follows\cite{Shaikh2019b,Xu2018,Hamil2023},
\begin{equation}
	x = \lim_{r \to \infty} \left( -r^2 \sin\theta_0 \frac{d\phi}{dr} \right)_{\theta_0 \to \frac{\pi}{2}},
	\quad
\end{equation}

\begin{equation}
	y = \lim_{r \to \infty} \left( r^2 \frac{d\theta}{dr} \right)_{\theta_0 \to \frac{\pi}{2}}.
\end{equation}
The celestial coordinate $x$ represents the observed perpendicular distance of the image, and $y$ represents the perpendicular distance projected onto the equatorial plane in Fig.\ref{Fig7} . The images clearly show that pressure suppresses the size of the shadow radius, while the modified gravity backgrounds with string cloud and quintessence enhance the size of the shadow radius.

\section{Thermodynamics and energy emission rate}
\label{emission}
In this section, we continue to study the thermodynamic structure of the black hole using physical observables. Besides the shadow radius, the maximum particle emission frequency is also an observable which can be directly measured. Next we focus on the connection between emission frequency and black hole thermodynamics, aiming to develop a new approach for probing black hole thermodynamics again. 

Some authors have shown that for a distant observer, the absorption cross-section tends to the size of the black hole shadow \cite{Wei2013,Belhaj2020arXiv1,Belhaj2020arXiv2}. In the scenario of extremely high energy, the absorption cross-section oscillates around the limiting value \(\sigma_{\text{lim}}\), which is expressed as
\begin{equation}
	\sigma_{\text{lim}} = \pi r_{s}^{2},
\end{equation}
where $r_{s}$ denotes the shadow radius of the black hole.

\begin{figure}[htbp]
	\begin{minipage}{0.325\linewidth}
		\centerline{\includegraphics[height=5cm,width=5cm]{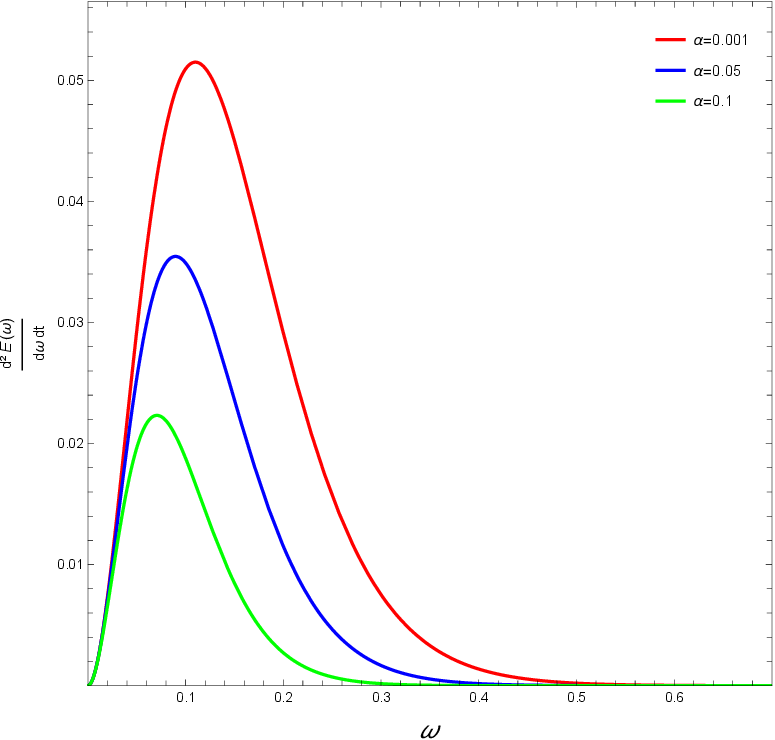}}
		\centerline{(a)}
	\end{minipage}
	\begin{minipage}{0.325\linewidth}
		\centerline{\includegraphics[height=5cm,width=5cm]{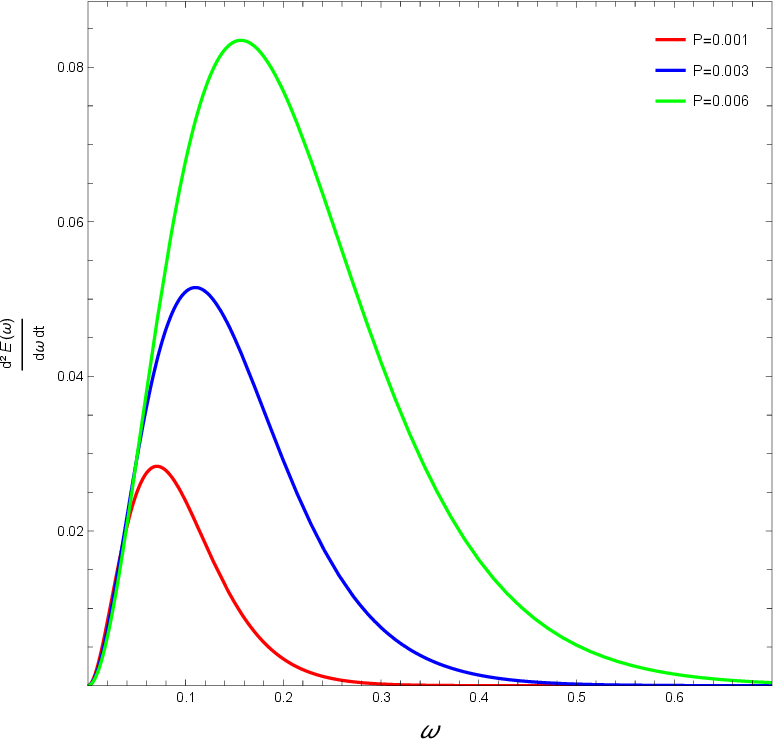}}
		\centerline{(b)}
	\end{minipage}
	\begin{minipage}{0.325\linewidth}
		\centerline{\includegraphics[height=5cm,width=5cm]{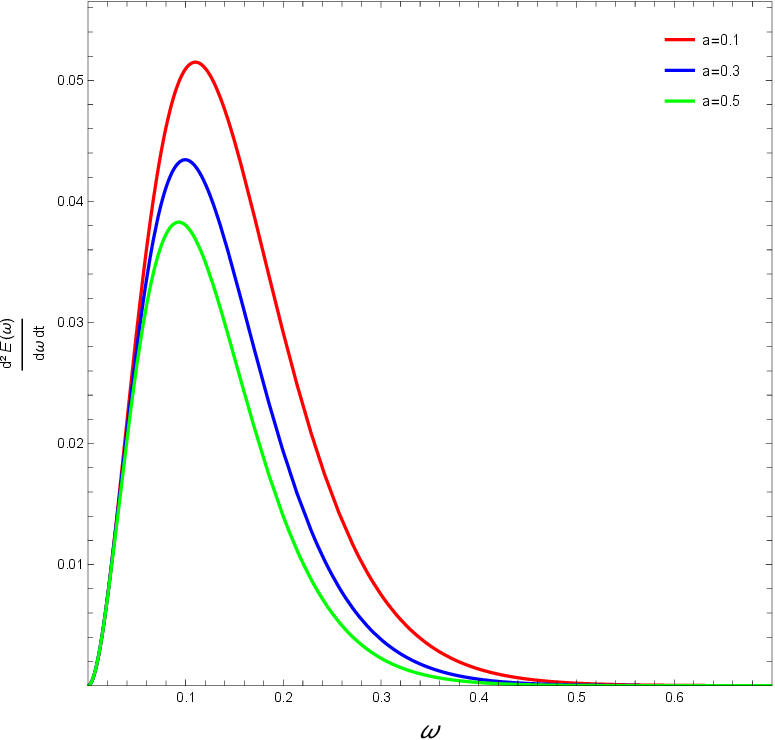}}
		\centerline{(c)}
	\end{minipage}
	\caption{The photon emission spectrum as a function of the photon emission frequency $\omega$ for the four-dimensional charged AdS black hole surrounded by string cloud and quintessence with different parameters.
		Figure $(a)$: Fixed the string cloud parameter $a=0.1$ and the thermodynamic pressure $P=0.003$, with varying quintessence parameter $\alpha$.
		Figure $(b)$: Fixed the string cloud parameter $a=0.1$ and the quintessence parameter $\alpha=0.001$, with varying thermodynamic pressure $P$.
		Figure $(c)$: Fixed the quintessence parameter $\alpha=0.001$ and the thermodynamic pressure $P=0.003$, with varying string cloud parameter $a$.
		Here we set $Q=1, r_0=100, M=2$.} \label{Fig8}
\end{figure}

The energy emission rate can be expressed as\cite{Panah2020,Ovgun2020}
\begin{equation}
\frac{d^2 E(\omega)}{d\omega dt} = \frac{2\pi \sigma_{\text{lim}}}{e^{\omega/T_H} - 1} \omega^3,
\end{equation}
where $\omega$ is the photon emission frequency, and $T_{H}$ is the Hawking temperature of the black hole. Figure.\ref{Fig8} shows the emission spectrum for different parameters. The figure clearly exhibits a distinct maximum peak in the emission spectrum, along with the corresponding maximum emission frequency $\omega_{\text{max}}$. The gravitational background from the string cloud and quintessence suppresses the overall amplitude of the emission spectrum and shifts its maximum emission frequency toward lower frequencies as the parameters increase. In contrast, an increase in the thermodynamic pressure enhances the amplitude of the black hole’s emission rate and shifts its peak toward higher frequencies. It is evident that gravitational corrections from the string cloud and quintessence slow down black hole evaporation. Meanwhile, since the thermodynamic pressure is related to the AdS radius which characterizes the natural curvature of the spacetime, the evaporation process is slow for a black hole located in a high curvature background.
   
\begin{figure}[htbp]
	\begin{minipage}{0.325\linewidth}
		\centerline{\includegraphics[height=5cm,width=5cm,keepaspectratio,trim=0 0 0 0,clip]{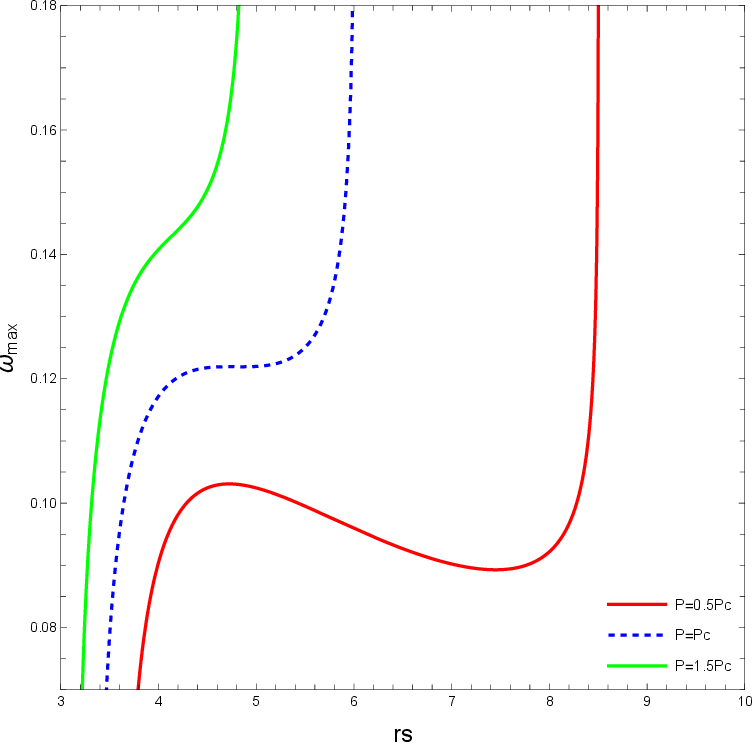}}
		\centerline{(a)}
	\end{minipage}
	\begin{minipage}{0.325\linewidth}
		\centerline{\includegraphics[height=5cm,width=5cm,keepaspectratio,trim=0 0 0 0,clip]{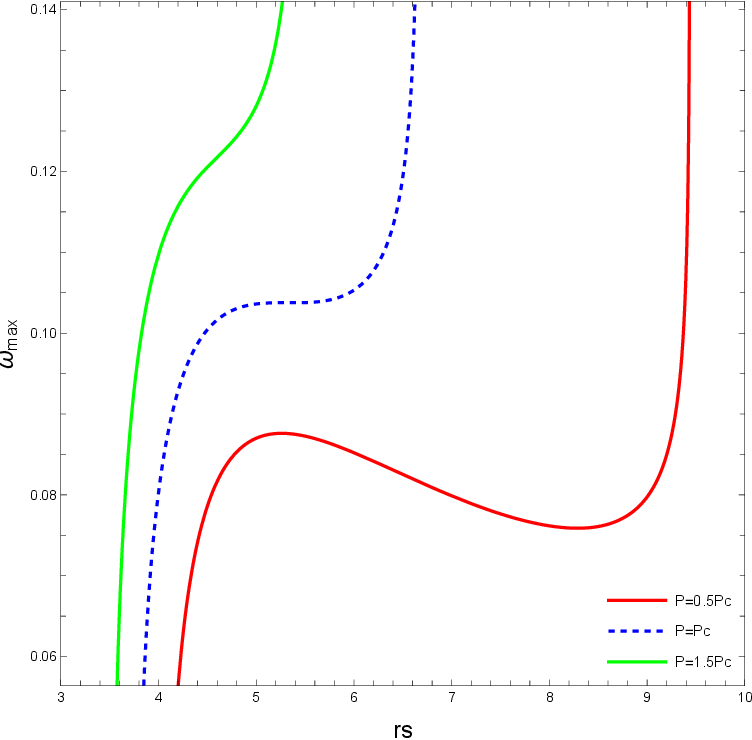}}
		\centerline{(b)}
	\end{minipage}
	\begin{minipage}{0.325\linewidth}
		\centerline{\includegraphics[height=5cm,width=5cm,keepaspectratio,trim=0 0 0 0,clip]{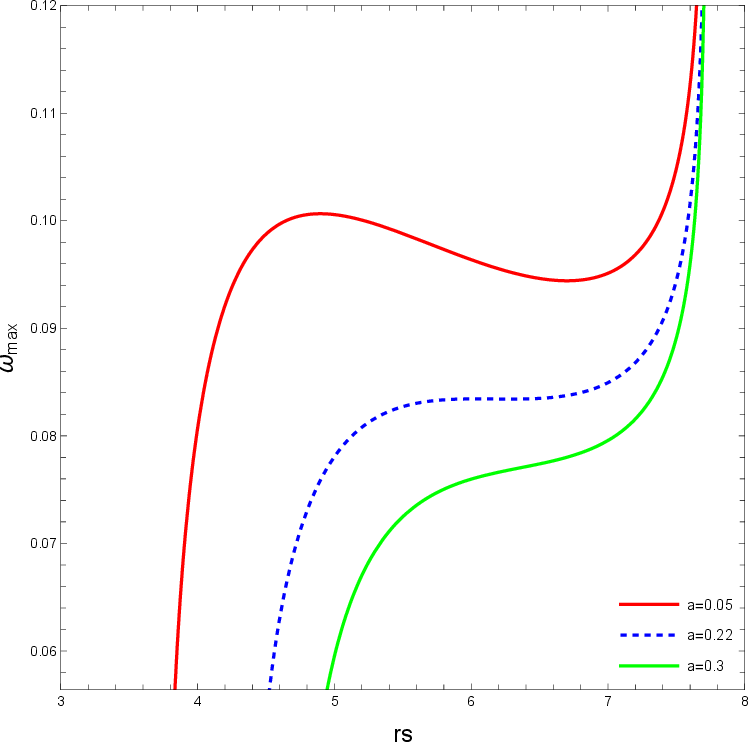}}
		\centerline{(c)}
	\end{minipage}
	\caption{The maximum frequency as a function of the shadow radius for the four-dimensional charged AdS black hole surrounded by strings cloud and quintessence with different parameters.
		Figure $(a)$: Fixed the string cloud parameter $a=0$ and the quintessence parameter $\alpha=0$, with varying thermodynamic pressure $P$.
		Figure $(b)$: Fixed the string cloud parameter $a=0.1$ and the quintessence parameter $\alpha=0.001$, with varying thermodynamic pressure $P$.
		Figure $(c)$: Fixed the thermodynamic pressure $P=0.002$ and the quintessence parameter $\alpha=0.001$, with varying string cloud parameter $a$.
		Here we set $Q=1, r_0=100$.} \label{Fig9}
\end{figure}

Now we solve for the maximum emission frequency $\omega_{\text{max}}$. The maximum emission frequency of the emission spectrum can be obtained from the following equation
\begin{equation}
	\frac{d}{d\omega}\left(\frac{d^2 E(\omega)}{d\omega dt}\right) = 0.
\end{equation}
we can directly derive Wien's displacement law, $\omega_{\text{max}} = 2.82\;T_{H}$. Inspired by the shadow formalism, the maximum emission frequency is also an observable quantity, allowing us to use \(\omega_{\text{max}}\) instead of \(T\) to study specific thermodynamic behaviors. Figure.\ref{Fig9} shows the relationship between the maximum emission frequency and the shadow radius. It is expected that the $\omega_{\text{max}}$–$r_{h}$ diagram reproduces the critical behavior observed in the $T$–$r_{h}$ plane. When \(P < P_c\), the maximum frequency exhibits local maximum and minimum, which indicates the occurrence of a small‑large black hole phase transition. At $P = P_c$, the function displays a single inflection point without any local extremum, marking the onset of critical behavior. Above the critical value, it becomes monotonic, indicating that the system remains in a stable phase. Changes in the string cloud background can also produce the results seen in the $T$-$r_h$ plane. The image results show that $\omega_{\text{max}}$ can also reveal phase transition behavior in black hole thermodynamics.

\begin{figure}[htbp]
	\begin{minipage}{0.325\linewidth}
		\centerline{\includegraphics[height=5cm,width=5cm,keepaspectratio,trim=0 0 0 0,clip]{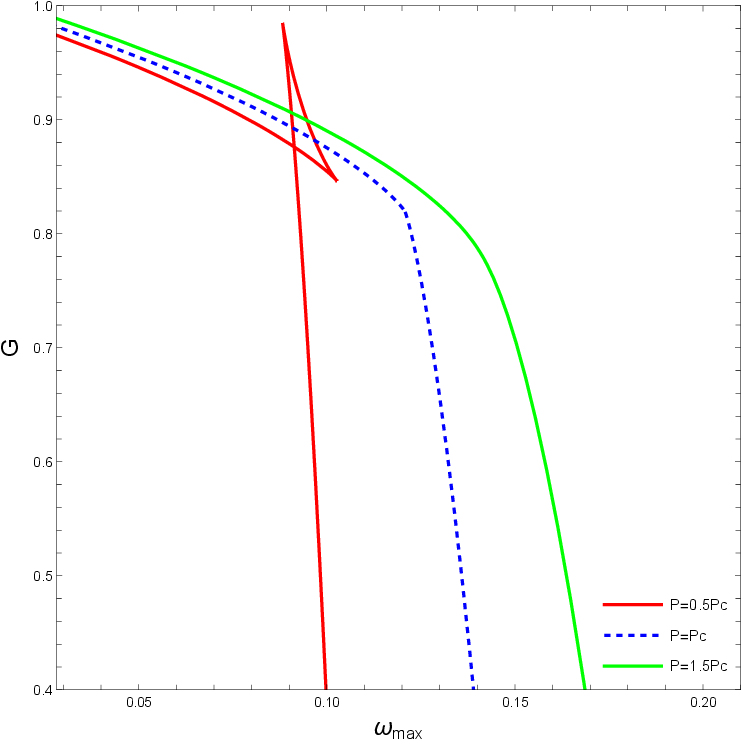}}
		\centerline{(a)}
	\end{minipage}
	\begin{minipage}{0.325\linewidth}
		\centerline{\includegraphics[height=5cm,width=5cm,keepaspectratio,trim=0 0 0 0,clip]{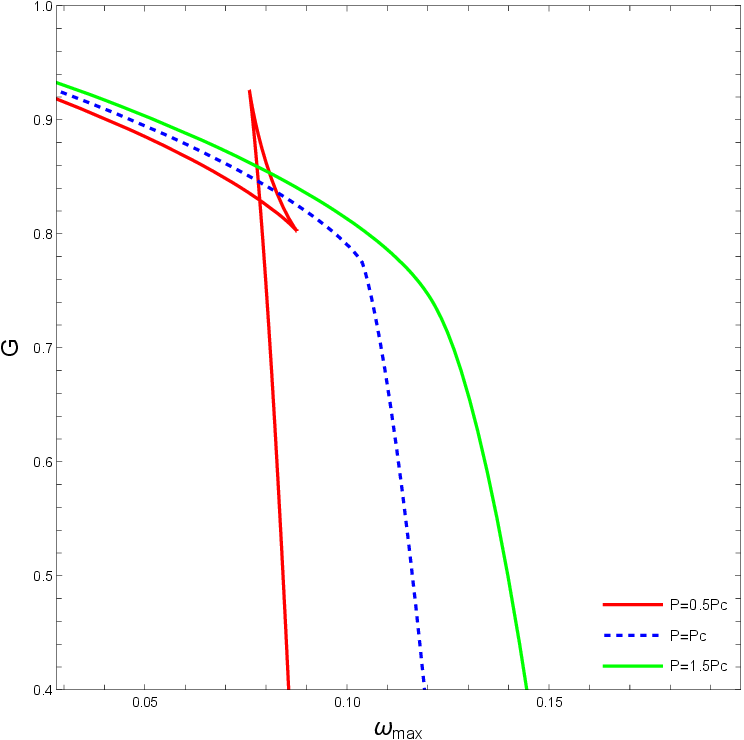}}
		\centerline{(b)}
	\end{minipage}
	\begin{minipage}{0.325\linewidth}
		\centerline{\includegraphics[height=5cm,width=5cm,keepaspectratio,trim=0 0 0 0,clip]{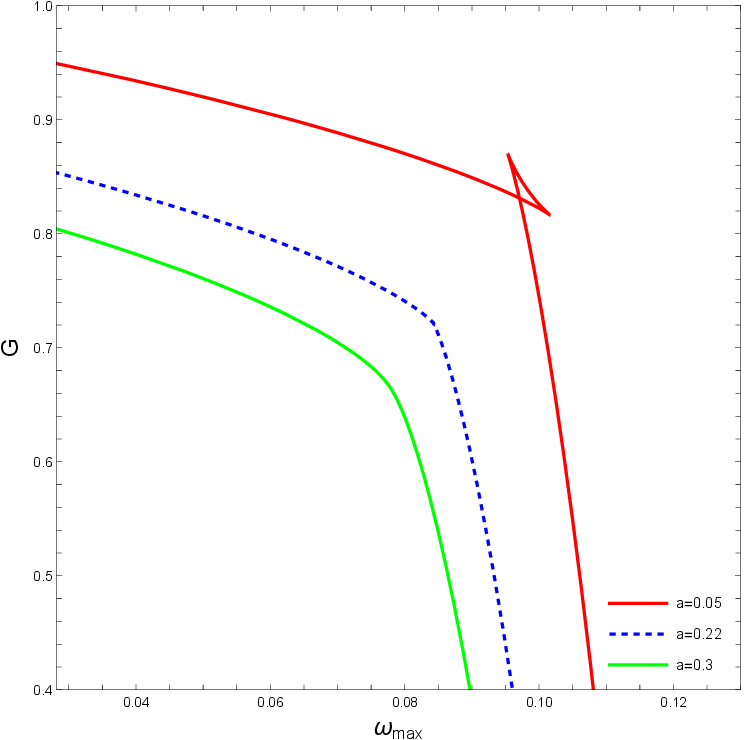}}
		\centerline{(c)}
	\end{minipage}
	\caption{The Gibbs free energy as a function of the maximum frequency for the four-dimensional charged AdS black hole surrounded by strings cloud and quintessence with different parameters.
		Figure $(a)$: Fixed the string cloud parameter $a=0$ and the quintessence parameter $\alpha=0$, with varying thermodynamic pressure $P$.
		Figure $(b)$: Fixed the string cloud parameter $a=0.1$ and the quintessence parameter $\alpha=0.001$, with varying thermodynamic pressure $P$.
		Figure $(c)$: Fixed the thermodynamic pressure $P=0.002$ and the quintessence parameter $\alpha=0.001$, with varying string cloud  parameter $a$.
		Here we set $Q=1, r_0=100$.} \label{Fig10}
\end{figure}

We plot the Gibbs free energy as a function of the maximum emission frequency in Fig.\ref{Fig10}. The swallowtail phase transition structure corresponding to the van der Waals fluid system reappears in the $G$-$T$ plane, which further confirms the maximum emission frequency of the emission spectrum is a reliable way to study the structure of black hole thermodynamic phase transitions.

\begin{figure}[htbp]
	\begin{minipage}{0.475\linewidth}
		\centerline{\includegraphics[height=4.8cm,width=6cm]{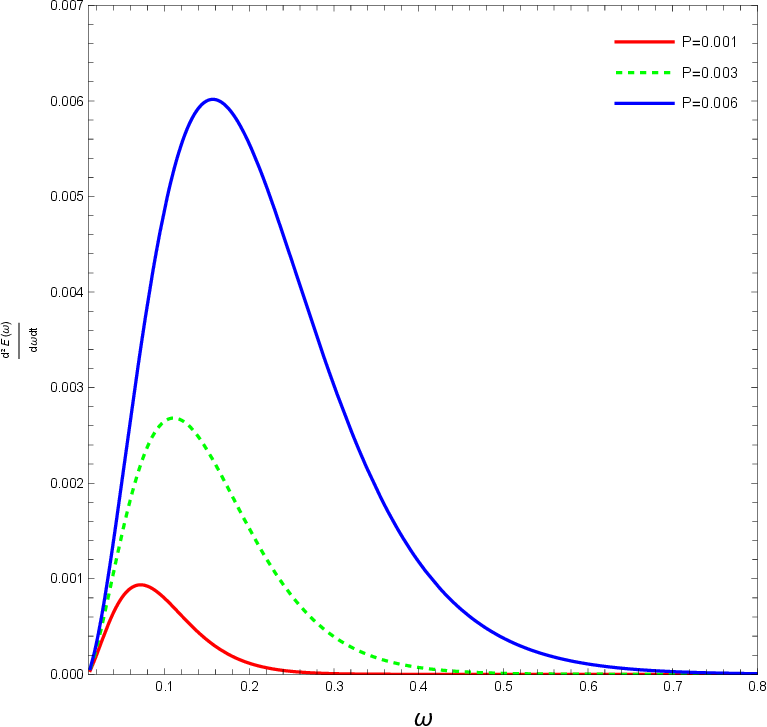}}
		\centerline{(a)}
	\end{minipage}
	\begin{minipage}{0.475\linewidth}
		\centerline{\includegraphics[height=4.8cm,width=6cm]{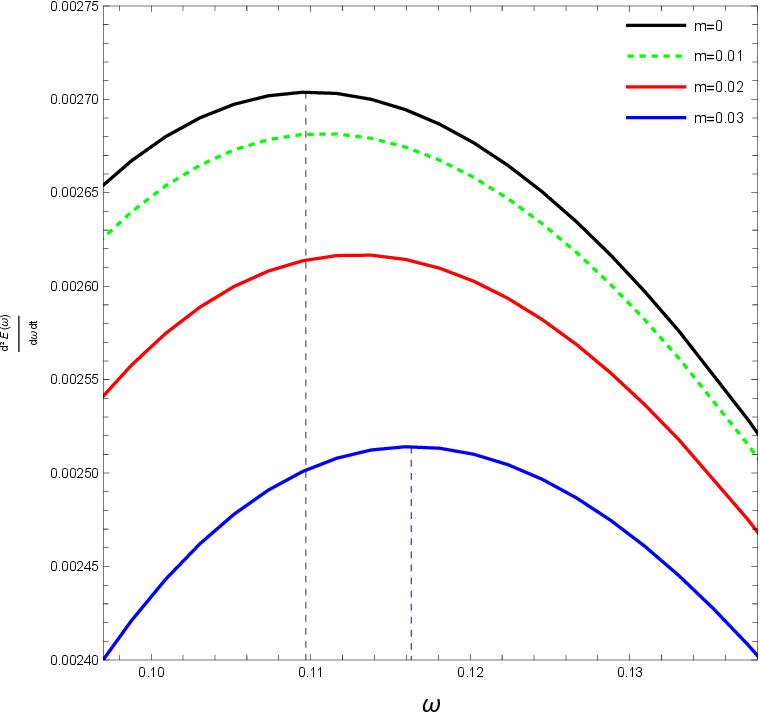}}
		\centerline{(b)}
	\end{minipage}
\caption{The emission rate of the massive field as a function of frequency for the four-dimensional charged AdS black hole surrounded by strings cloud and quintessence with different parameters.
	Figure $(a)$: Fixed the cloud of string parameter $a=0.1$, the quintessence parameter $\alpha=0.001$ and the particle mass $m=0.01$, with varying thermodynamic pressure $P$.
	Figure $(b)$: Fixed the cloud of string parameter $a=0.1$ and the quintessence parameter $\alpha=0.001$, with varying particle mass $m$.
	Here we set $Q=1, r_0=100, M=2$.}
	\label{Fig11}
\end{figure}

After discussing photon emission around black holes, we analyze the emission behavior of massive particles around them. Since a black hole absorbs massive particles, we take the horizon area as an approximation for the absorption cross section. The absorption cross section is approximated as
\begin{equation}
A_{\text{lim}} = \pi r_h^2.
\end{equation}
The emission rate for massive particles is given by \cite{Emparan2000},
\begin{equation}
\frac{d^2 E(\omega, m)}{d\omega dt} \simeq \left(\omega^2 - m^2\right) \omega \frac{A_{\text{lim}}}{e^{\omega/T_H} - 1}.
\end{equation}
In Fig.\ref{Fig11} , we first plot the emission spectrum of the massive field for different pressures and select the spectrum at the critical pressure to analyze the influence of different particle masses. The emission rate of massive particles behaves the same as that of massless particles for different pressures. We observe the following phenomena for massive particle case: As the particle mass $m$ increases, the maximum emission frequency shifts to the right and the overall amplitude of the emission rate decreases. A larger emission amplitude implies easier observability, i.e. lower-mass particles are more readily detectable. It is important to measure the quantum black hole in the highest-energy particle colliders, because their production and decay signatures would be dominated by processes involving lighter fundamental particles.

The presence of massive particles introduces new shifts in the maximum emission frequency, leading to a generalized Wien’s displacement law
 \begin{equation}
 	\omega_{\text{max}}(m, T_H) \approx 0.60\, T_H + \frac{0.28\, m^2 + 1.11\, T_H^2}{\left(0.37\, m^3 + 0.94\, T_H^3\right)^{1/3}} + 1.11\,\left(0.37\, m^3 + 0.94\, T_H^3\right)^{1/3}.
 	\label{eq:omega_max}
 \end{equation}
 When $m=0$, the classical Wien's law is recovered and $\omega_{\text{max}}$ returns to $2.82T_H$.
 
\begin{figure}[htbp]
	\begin{minipage}{0.475\linewidth}
		\centerline{\includegraphics[height=4.8cm,width=6cm]{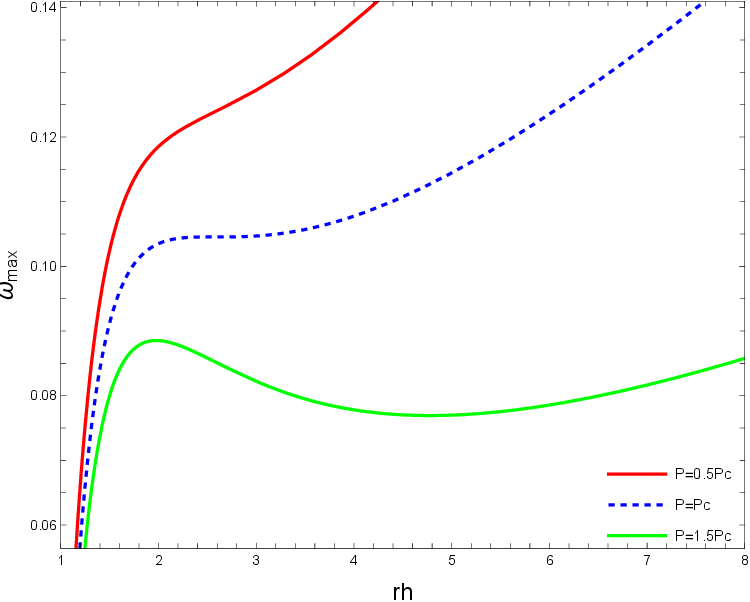}}
		\centerline{(a)}
	\end{minipage}
	\begin{minipage}{0.475\linewidth}
		\centerline{\includegraphics[height=4.8cm,width=6cm]{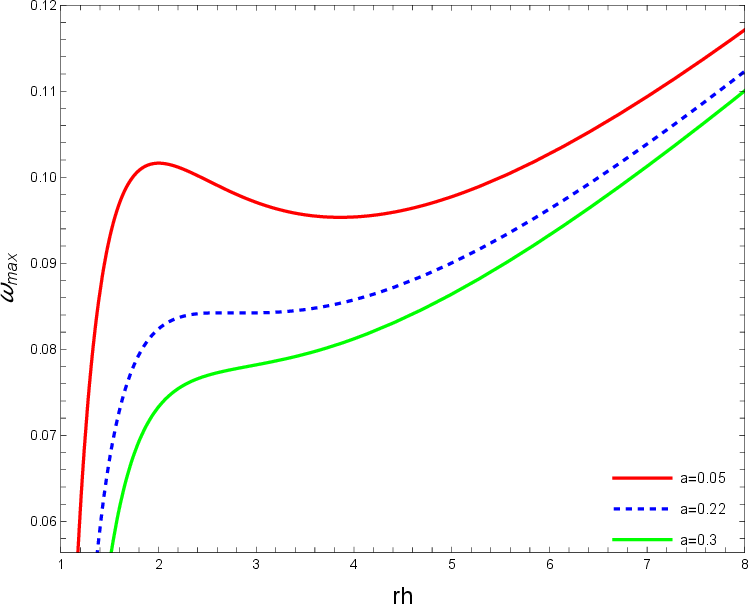}}
		\centerline{(b)}
	\end{minipage}
	\caption{The maximum emission frequency as a function of the event horizon radius in the massive field for the four-dimensional charged AdS black hole surrounded by string cloud and quintessence with different parameters.
		Figure $(a)$: Fixed the string cloud parameter $a=0.1$ and the quintessence parameter $\alpha=0.001$, with varying thermodynamic pressure $P$.
		Figure $(b)$: Fixed the thermodynamic pressure $P=0.002$ and the quintessence parameter $\alpha=0.001$, with varying string cloud parameter $a$.
		Here we set $Q=1, r_0=100$.}
	\label{Fig12}
\end{figure}

\begin{figure}[htbp]
	\begin{minipage}{0.475\linewidth}
		\centerline{\includegraphics[height=4.8cm,width=6cm]{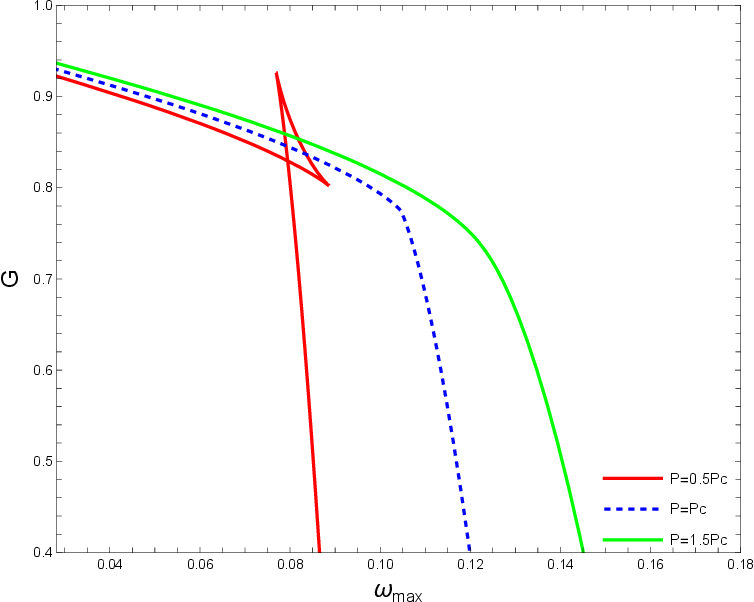}}
		\centerline{(a)}
	\end{minipage}
	\begin{minipage}{0.475\linewidth}
		\centerline{\includegraphics[height=4.8cm,width=6cm]{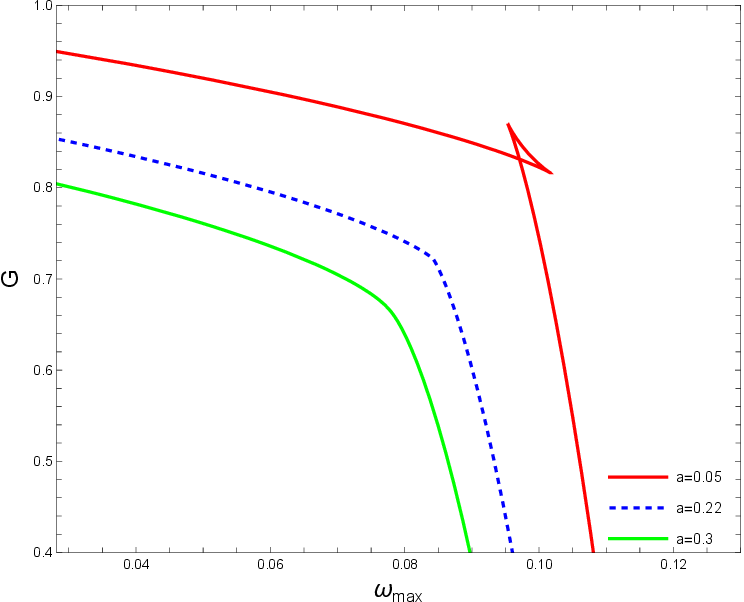}}
		\centerline{(b)}
	\end{minipage}
	\caption{The Gibbs free energy as a function of the maximum emission frequency in the massive field for the four-dimensional charged AdS black hole surrounded by string cloud and quintessence with different parameters.
		Figure $(a)$: Fixed the string cloud parameter $a=0.1$ and the quintessence parameter $\alpha=0.001$, with varying thermodynamic pressure $P$.
		Figure $(b)$: Fixed the thermodynamic pressure $P=0.002$ and the quintessence parameter $\alpha=0.001$, with varying string cloud parameter $a$.
		Here we set $Q=1, r_0=100$.}
	\label{Fig13}
\end{figure}

Finally, we plot the variation of the maximum emission frequency $\omega_{\text{max}}$ with the event horizon radius $r_h$ in the massive field (Fig.~\ref{Fig12}), and the Gibbs free energy $G$ versus the maximum emission frequency $\omega_{\text{max}}$ (Fig.~\ref{Fig13}). Similar to the case of photons, the phase transition structure of black hole thermodynamics appears fully again. The results show that using the emission spectrum of massive particles to probe black hole thermodynamics is also feasible, especially for micro black holes in high-energy particle collisions.

Within physically viable parameter spaces, both the horizon radius $r_h$ and shadow radius $r_s$ solved from the model are same order unity in natural units. The dimensionless framework describes supermassive black holes hosted at galactic centers, which are key targets for high-resolution astronomical observations carried out by the Event Horizon Telescope (EHT). Black hole shadows, as direct observables, have been measured in images of M87* and Sgr A*, which confirmed their practical observational relevance for the study of supermassive black holes.

Though current instruments cannot directly detect Hawking radiation emitted by supermassive black holes, the shadow radius and peak radiation frequency studied here are physically meaningful real observables. The string cloud and quintessence background fields produce systematic deviations in shadow size and peak radiation frequency, which are theoretically distinguishable and can serve as effective probes of exotic cosmic backgrounds.

While experimental verification requires high-precision observational facilities, the physical quantities and phase transition behaviors discussed here exhibit clear correspondence with observational data. The findings provide concrete theoretical support for future tests of gravitational theories and dark background models using black hole shadow and radiation measurements.

\section{Conclusion}
\label{conclusion}
In this paper, we have investigated the shadow thermodynamics and particle emission spectrum of the four‑dimensional charged AdS black hole surrounded by string cloud and quintessence. We first have established the photon orbit in the equatorial plane of the black hole using the Euler-Lagrangian equation. Then, we have obtained the equation for the black hole shadow radius, where the background of the string cloud and quintessence causes an increase of shadow radius, however, the thermodynamic pressure exerts an opposite influence by suppressing the shadow radius. Our analysis has showed that the shadow radius and the event horizon radius have a positive correlation for different parameters. We have verified the monotonic and invertible mapping between the black hole shadow radius and the horizon radius, and ensured the universality of the observable proxy method. By analyzing temperature and heat capacity of the black hole, it can be seen that the thermodynamic behavior of black holes is perfectly replicated in the analysis with shadow radius as a variable, which exhibits a phase structure similar to that of a van der Waals fluid. The observability of the black hole shadow makes it easier for us to observe these thermodynamic features of black holes.

Additionally, we have investigated the particle emission rate around the black hole through the shadow radius. We have found that gravitational corrections from the string cloud and quintessence slow down black hole evaporation. Meanwhile, a high curvature background also leads to slower evaporation because it corresponds to a smaller AdS radius and higher thermodynamic pressure. Because the peak frequency is sensitive to the background field it can be used as a new probe to identify the dark environment around a black hole. A redshift or blueshift of the peak frequency directly encodes the information from string cloud quintessence and thermodynamic pressure. Because the maximum emission frequency $\omega_{\text{max}}$ of the emission spectrum is physical observable, we can use $\omega_{\text{max}}$ to obtain an image that shows behavior similar to a thermodynamic phase transition. By taking the derivative of the emission rate formula, we have found that the maximum emission frequency $\omega_{\text{max}}$ also has a positive linear relationship with the temperature $T$, which allows us to use $\omega_{\text{max}}$ as a substitute for temperature $T$. The results have showed that our approach is valid, and the $\omega_{\text{max}}$-$r_h$ plane reproduced the behavior of the $T$-$r_h$ plane. The same phase transition structure also appears in the $\omega_{\text{max}}$-$r_s$ plane. 

Finally, massive particles cause new shifts in the maximum emission frequency, which leads to a generalized Wien’s displacement law. Specifically, $\omega_{\text{max}}$ shifts to the right and the emission spectrum magnitude decreases as the particle mass increases. If quantum black holes are detected in the highest-energy particle colliders in the future, the detection of light particles will be especially important, and our finding shows that light particles are easier to detect.

\begin{acknowledgments}
This work is supported by the National Science Foundation of China under Grant Nos. 12373022, U1731107.
\end{acknowledgments}


\begin{thebibliography}{00}

\bibitem{abbott2016prl3} Abbott B P et al., {\sl Phys. Rev. Lett.} {\bf116}, 061102 (2016).

\bibitem{abbott2016prl2} Abbott B P et al., {\sl Phys. Rev. Lett.} {\bf116}, 241102 (2016).

\bibitem{abbott2016prl1} Abbott B P et al., {\sl Phys. Rev. Lett.} {\bf116}, 241103 (2016).

\bibitem{AK2019L1} Akiyama K et al., {\sl Astrophys. J. Lett.} {\bf875}, L1-L6 (2019).

\bibitem{akiyama2022ajl12} Akiyama K et al., {\sl Astrophys. J. Lett.} {\bf930}, L12-L16 (2022).

\bibitem{Vazquez2004Lensing} Vazquez S E and Esteban E P, {\sl Nuovo Cim. B} {\bf119}, 489–519 (2004).

\bibitem{Shaikh2019Shadows} Shaikh R, Kocherlakota P, Narayan R, and Joshi P S, {\sl Mon. Not. Roy. Astron. Soc.} {\bf482}, 52 (2019).

\bibitem{Hou2018Shadow} Hou X, Xu Z, Zhou M, and Wang J, {\sl JCAP} {\bf1807} no. 07, 015 (2018).

\bibitem{Cunha2018Shadow} Cunha P V P, Herdeiro C A R, and Rodriguez M J, {\sl Phys. Rev. D} {\bf97} no. 8, 084020 (2018).

\bibitem{Tsukamoto2018Shadow} Tsukamoto N, {\sl Phys. Rev. D} {\bf97} no. 6, 064021 (2018).

\bibitem{Wald1984GeneralRelativity} Wald R M, {\sl General Relativity} (The University of Chicago Press, 1984).

\bibitem{Bardeen1973} Bardeen J M, {\sl in black holes. In: Proceedings, Ecole d'Ete de Physique Theorique: Les Astres Occlus, Les Houches, France, August 1972}, pp. 215–240 (1973).

\bibitem{Chandrasekhar1983} Chandrasekhar S, {\sl The Mathematical Theory of Black Holes} (Oxford: Oxford University Press, 1984).

\bibitem{Bambi2009} Bambi C and Freese K, {\sl Phys. Rev. D} {\bf79}, 043002 (2009).

\bibitem{Bambi2010} Bambi C and Yoshida N, {\sl Class. Quant. Grav.} {\bf27}, 205006 (2010).

\bibitem{Atamurotov2013} Atamurotov F, Abdujabbarov A and Ahmedov B, {\sl Phys. Rev. D} {\bf88} no. 6, 064004 (2013).

\bibitem{Papnoi2014} Papnoi U, Atamurotov F, Ghosh S G and Ahmedov B, {\sl Phys. Rev. D} {\bf90} no. 2, 024073 (2014).

\bibitem{Atamurotov2015} Atamurotov F and Ahmedov B, {\sl Phys. Rev. D} {\bf92}, 084005 (2015).

\bibitem{Wang2018} Wang M, Chen S and Jing J, {\sl Phys. Rev. D} {\bf97} no. 6, 064029 (2018).

\bibitem{Guo2018} Guo M, Obers N A and Yan H, {\sl Phys. Rev. D} {\bf98} no. 8, 084063 (2018).

\bibitem{Yan2019} Yan H, {\sl Phys. Rev. D} {\bf99} no. 8, 084050 (2019).

\bibitem{Konoplya2019} Konoplya R A, {\sl Phys. Lett. B} {\bf795}, 1-6 (2019).

\bibitem{Perlick2022} Perlick V and Tsupko O Y, {\sl Phys. Rept.} {\bf947}, 1 (2022).

\bibitem{Synge1966b} Synge J L, {\sl Mon. Not. R. Astron. Soc.} {\bf131} no. 3, 463-466 (1966).

\bibitem{Amarilla2010} Amarilla L, Eiroa E F and Giribet G, {\sl Phys. Rev. D} {\bf81}, 124045 (2010).

\bibitem{Luminet1979b} Luminet J P, {\sl Astron. Astrophys.} {\bf75}, 228-235 (1979).

\bibitem{Afrin2021} Afrin M, Kumar R and Ghosh S G, {\sl Mon. Not. Roy. Astron. Soc.} {\bf504}, 5927-5940 (2021).

\bibitem{Khodadi2021} Khodadi M, Lambiase G, Mota D F, {\sl J. Cosmol. Astropart. Phys.} {\bf2021}, 028 (2021).

\bibitem{Ghosh2022} Ghosh S G, Afrin M, {\sl Phys. Rev. D} {\bf106}, 024029 (2022).

\bibitem{Cunha2022} Cunha P V P, Herdeiro C A R, Radu E, {\sl Phys. Rev. D} {\bf106}, 024029 (2022).

\bibitem{Hawking1975} Hawking S W, {\sl Commun. Math. Phys.} {\bf43}, 199-220 (1975).

\bibitem{HawkingPage1983} Hawking S W and Page D N, {\sl Commun. Math. Phys.} {\bf87}, 577 (1983).

\bibitem{Hawking1974} Hawking S W, {\sl Nature} {\bf248}, 30-31 (1974).

\bibitem{Bekenstein1973a} Bekenstein J D, {\sl Phys. Rev. D} {\bf7}, 2333 (1973).

\bibitem{Bekenstein1972} Bekenstein J D, {\sl Lettere al Nuovo Cimento} {\bf4}, 737 (1972).

\bibitem{Bekenstein1973b} Bekenstein J D, {\sl Phys. Rev. D} {\bf7}, 949 (1973).

\bibitem{Bekenstein1974} Bekenstein J D, {\sl Phys. Rev. D} {\bf9}, 3292 (1974).

\bibitem{Bekenstein1975} Bekenstein J D, {\sl Phys. Rev. D} {\bf12}, 3077 (1975).

\bibitem{Bardeen1973FourLaws} Bardeen J M, Carter B and Hawking S W, {\sl Commun. Math. Phys.} {\bf31}, 161 (1973).

\bibitem{Kubiznak2012} Kubiznak D and Mann R B, {\sl JHEP} {\bf07}, 033 (2012).

\bibitem{Cai2013} Cai R G, Cao L M, Li L and Yang R Q, {\sl JHEP} {\bf09}, 005 (2013).

\bibitem{Kastor2009} Kastor D, Ray S and Traschen J, {\sl Class. Quant. Grav.} {\bf26}, 195011 (2009).

\bibitem{Dolan2011} Dolan B P, {\sl Class. Quantum Grav.} {\bf28}, 235017 (2011).

\bibitem{Page1976} Page D N, {\sl Phys. Rev. D} {\bf13}, 198-206 (1976).

\bibitem{Emparan2000} Emparan R, Horowitz G T and Myers R C, {\sl Phys. Rev. Lett.} {\bf85}, 499-502 (2000).

\bibitem{Magueijo2004} Magueijo J, Smolin L, {\sl Classical Quantum Gravity} {\bf21}, 1725-1736 (2004).

\bibitem{deRham2010} de Rham C, Gabadadze G, {\sl Phys. Rev. D} {\bf82}, 044020 (2010).

\bibitem{deRham2011} de Rham C, Gabadadze G and Tolley A J, {\sl Phys. Rev. Lett.} {\bf106}, 231101 (2011).

\bibitem{Letelier1979} Letelier P S, {\sl Phys. Rev. D} {\bf20} no. 6, 1294 (1979).

\bibitem{Perlmutter1999} Perlmutter S, Aldering G, Goldhaber G, Knop R, Nugent P, Castro P, Deustua S, Fabbro S, Goobar A, Groom D E et al., {\sl Astrophys. J.} {\bf517} no. 2, 565 (1999).

\bibitem{Kiselev2003} Kiselev V V, {\sl Class. Quant. Grav.} {\bf20} no. 6, 1187 (2003).

\bibitem{Maldacena1998} Maldacena J M, {\sl Adv. Theor. Math. Phys.} {\bf2}, 231-252 (1998).

\bibitem{Witten1998} Witten E, {\sl Adv. Theor. Math. Phys.} {\bf2}, 253-291 (1998).

\bibitem{Gubser1998} Gubser S S, Klebanov I R and Polyakov A M, {\sl Phys. Lett. B} {\bf428}, 105-114 (1998).

\bibitem{Cvetic2011} Cvetic M, Gibbons G W, Kubiznak D and Pope C N, {\sl Phys. Rev. D} {\bf84}, 024037 (2011).

\bibitem{Chamblin1999a} Chamblin A, Emparan R, Johnson C V and Myers R C, {\sl Phys. Rev. D} {\bf60}, 064018 (1999).

\bibitem{Chamblin1999b} Chamblin A, Emparan R, Johnson C V and Myers R C, {\sl Phys. Rev. D} {\bf60}, 104026 (1999).

\bibitem{Caldarelli2000} Caldarelli M M, Cognola G and Klemm D, {\sl Class. Quant. Grav.} {\bf17}, 399-420 (2000).

\bibitem{Toledo2019} Toledo J M and Bezerra V B, {\sl Int. J. Mod. Phys. D} {\bf28}, 1950023 (2019).

\bibitem{Li2023} Li X Q, Yan H P, Xing L L and Zhou S W, {\sl Phys. Rev. D} {\bf107} no. 10, 104055 (2023).

\bibitem{Zhang2020} Zhang M and Guo M, {\sl Eur. Phys. J. C} {\bf80} no. 8, 790 (2020).

\bibitem{Belhaj2020} Belhaj A, Chakhchi L, El Moumni H, Khalloufi J and Masmar K, {\sl Int. J. Mod. Phys. A} {\bf35} no. 27, 2050170 (2020).

\bibitem{Cai2021arXiv} Cai X C and Miao Y G, {\sl arXiv:2107.08352 [gr-qc]} (2021).

\bibitem{Wang2022} Wang C, Wu B, Xu Z M and Yang W L, {\sl Nucl. Phys. B} {\bf976}, 115698 (2022).

\bibitem{Shaikh2019b} Shaikh R, {\sl Phys. Rev. D} {\bf100} no. 2, 024028 (2019).

\bibitem{Xu2018} Xu Z, Hou X and Wang J, {\sl JCAP} {\bf10}, 046 (2018).

\bibitem{Hamil2023} Hamil B and Lu¨tfu¨o˘glu B C, {\sl Phys. Dark Univ.} {\bf42}, 101293 (2023).

\bibitem{Wei2013} Wei S W and Liu Y X, {\sl J. Cosmol. Astrop. Phys.} {\bf11}, 063 (2013).

\bibitem{Belhaj2020arXiv1} Belhaj A, Benali M, El Balali A, El Moumni H and Ennadifi S E, {\sl arXiv:2006.01078} (2020).

\bibitem{Belhaj2020arXiv2} Belhaj A, Benali M, El Balali A, El Hadri W and El Moumni H, {\sl arXiv:2007.09058} (2020).

\bibitem{Panah2020} Eslam Panah B, Jafarzade K and Hendi S H, {\sl Nucl. Phys. B} {\bf961}, 115269 (2020).

\bibitem{Ovgun2020} O¨ vgu¨n A, Sakallı ˙I, Saavedra J and Leiva C, {\sl Mod. Phys. Lett. A} {\bf35} no. 20, 2050163 (2020).
\end{thebibliography}
\end{document}